\documentclass[twocolumn]{aastex631}

\usepackage{amsmath}
\let\oldAA\AA
\renewcommand{\AA}{\text{\normalfont\oldAA}}

\received{2022 Nov.2}
\revised{2023 Aug.22}
\accepted{2023 Aug.28}
\submitjournal{ApJL}

\shorttitle{MZR of Dwarfs at $z=2-3$}
\shortauthors{M.Li et al.}
\graphicspath{{./}{figures/}}

\begin{document}
\title{The Mass-Metallicity Relation of Dwarf Galaxies at Cosmic Noon from JWST Observations}

\author[0000-0001-6251-649X]{Mingyu Li}
\affiliation{Department of Astronomy, Tsinghua University, Beijing 100084, China\email{lmytime@hotmail.com, zcai@mail.tsinghua.edu.cn}}

\author[0000-0001-8467-6478]{Zheng Cai}
\affiliation{Department of Astronomy, Tsinghua University, Beijing 100084, China\email{lmytime@hotmail.com, zcai@mail.tsinghua.edu.cn}}

\author[0000-0002-1620-0897]{Fuyan Bian}
\affiliation{European Southern Observatory, Alonso de C\'{o}rdova 3107, Casilla 19001, Vitacura, Santiago 19, Chile\email{Fuyan.Bian@eso.org}}

\author[0000-0001-6052-4234]{Xiaojing Lin}
\affiliation{Department of Astronomy, Tsinghua University, Beijing 100084, China\email{lmytime@hotmail.com, zcai@mail.tsinghua.edu.cn}}

\author[0000-0001-5951-459X]{Zihao Li}
\affiliation{Department of Astronomy, Tsinghua University, Beijing 100084, China\email{lmytime@hotmail.com, zcai@mail.tsinghua.edu.cn}}

\author[0000-0003-0111-8249]{Yunjing Wu}
\affiliation{Department of Astronomy, Tsinghua University, Beijing 100084, China\email{lmytime@hotmail.com, zcai@mail.tsinghua.edu.cn}}
\affiliation{Steward Observatory, University of Arizona, 933 N Cherry Ave, Tucson, AZ 85721, USA}

\author[0000-0002-4622-6617]{Fengwu Sun}
\affiliation{Steward Observatory, University of Arizona, 933 N Cherry Ave, Tucson, AZ 85721, USA}

\author[0000-0002-0427-9577]{Shiwu Zhang}
\affiliation{Department of Astronomy, Tsinghua University, Beijing 100084, China\email{lmytime@hotmail.com, zcai@mail.tsinghua.edu.cn}}

\author[0000-0001-5160-6713]{Emmet Golden-Marx}
\affiliation{Department of Astronomy, Tsinghua University, Beijing 100084, China\email{lmytime@hotmail.com, zcai@mail.tsinghua.edu.cn}}

\author[0000-0002-8246-7792]{Zechang Sun}
\affiliation{Department of Astronomy, Tsinghua University, Beijing 100084, China\email{lmytime@hotmail.com, zcai@mail.tsinghua.edu.cn}}

\author[0000-0002-3983-6484]{Siwei Zou}
\affiliation{Department of Astronomy, Tsinghua University, Beijing 100084, China\email{lmytime@hotmail.com, zcai@mail.tsinghua.edu.cn}}

\author[0000-0003-3310-0131]{Xiaohui Fan}
\affil{Steward Observatory, University of Arizona, 933 N Cherry Ave, Tucson, AZ 85721, USA}

\author[0000-0003-1344-9475]{Eiichi Egami}
\affiliation{Steward Observatory, University of Arizona, 933 N Cherry Ave, Tucson, AZ 85721, USA}

\author[0000-0003-3458-2275]{Stephane Charlot}
\affiliation{Sorbonne Universit\'e, CNRS, UMR7095, Institut d'Astrophysique de Paris, F-75014, Paris, France}

\author[0000-0002-6971-5755]{Gustavo Bruzual}
\affiliation{Institute of Radio Astronomy and Astrophysics, National Autonomous University of Mexico, San José de la Huerta
58089 Morelia, Michoacán, México}

\author[0000-0002-7636-0534]{Jacopo Chevallard}
\affiliation{Department of Physics, University of Oxford, Denys Wilkinson Building, Keble Road, Oxford OX1 3RH, UK}

\begin{abstract}
We present a study of the mass-metallicity relation (MZR) of 51 dwarf galaxies
($M_\star\approx 10^{6.5} - 10^{9.5}\mathrm{M}_\odot$) at $z = 2-3$ from the Abell 2744 and SMACS J0723-3732 galaxy cluster fields.
These dwarf galaxies are identified and confirmed by deep JWST/NIRISS imaging and slitless grism spectroscopic observations.
By taking advantage of the superior performance of JWST and the gravitational lensing effect, we extend the previous MZR relation at $z=2-3$ to a much lower mass regime down by $\approx$ 2.5 orders of magnitude as compared with previous studies.
We find that the MZR has a shallower slope at the low-mass end ($M_\star<10^{9}~\rm \mathrm{M}_\odot$), with a slope turnover point of $\approx$ $10^9~\rm  \mathrm{M}_\odot$.
This implies that the dominating feedback processes in dwarf galaxies may be different from that in massive galaxies.
From $z=3$ to $z=2$, the metallicity of the dwarf galaxies is enhanced by $\approx0.09$ dex for a given stellar mass, consistent with the mild evolution found in galaxies with higher mass.
Furthermore, we confirm the existence of a fundamental metallicity relation (FMR) between the gas-phase metallicity, stellar mass, and star formation rate in dwarf galaxies at $z=2-3$.
Our derived FMR, which has no significant redshift evolution, can be used as a benchmark to understand the origin of the anti-correlation between the SFR and metallicity of dwarf galaxies in the high-$z$ Universe.
\end{abstract}


\section{Introduction}\label{sec:intro}
The mass-metallicity relation (MZR) is one of the most fundamental relations {for quantifying} the chemical enrichment and gas properties {of} star-forming galaxies {and} has been studied across cosmic time
\citep[e.g.,][]{Tremonti2004, Erb2006, Maier2014, Steidel2014, Sanders2015, Guo2016, Sanders2021, Papovich2022, Li2022}.
In this scaling relation, the metallicity increases with stellar mass, and galaxies at high redshifts tend to have lower metallicity than their low-redshift counterparts for a given stellar mass.
The normalization, slope, and scatter of the MZR provide insight into the physical mechanisms regulating galaxy evolution, including how metals have been produced and driven out of galaxies and how pristine/metal-populated gas has been accreted into galaxies \citep[e.g.,][]{Peeples2011, Lilly2013,Torrey2019}.

Dwarf galaxies are the most numerous type of observed galaxies.
They are believed to dominate the ionizing photon budget and play the most important role in the metal enrichment of the circum-/intergalacitic medium (CGM/IGM) \citep[e.g.,][]{Robertson2010,Lin2022,Wu2021}, giving them great cosmological importance. 
Understanding the gas properties of dwarf galaxies and studying the evolution of gas-phase metallicity of dwarf galaxies at a given stellar mass can provide insight into the baryonic processes that regulate star formation and galaxy growth throughout cosmic history.
Nevertheless, to date, most studies focus on the MZR of high to intermediate mass galaxies ($>$ 10$^9$ $\mathrm{M}_\odot$). 
At these mass ranges, there have been multiple examples of the MZR 
at different redshifts {haing} similar slopes \citep[e.g.,][]{Zahid2014b, Sanders2021}. 

At low redshift, \citet{Zahid2012} studied the MZR $z\sim0$ using a sample of local dwarf galaxies.
They found that the scatter of MZR increases with decreasing stellar mass. 
Similarly, \citet{Guo2016} identified an increase in MZR scatter in a sample of dwarf galaxies at $0.5<z<0.7$ and tentatively suggested a shallower MZR slope when the 
stellar mass is lower than $10^9 ~\mathrm{M_\odot}$.
However, the MZR has been probed in very few high-$z$ ($z > 2$), low-mass galaxies ($<$ $10^9\mathrm{M}_\odot$).
Therefore, it is unclear whether the MZR has the same slope at the low mass end and how the MZR for dwarf galaxies evolves with cosmic time beyond $z=2$. 

In this paper, we probe the MZR in a sample of 51 dwarf galaxies ($M_\star\approx 10^{6.5}-10^{9.5}\ \mathrm{M}_\odot$) at $z = 2-3$ using Near-Infrared Imager and Slitless Spectrograph (NIRISS) grism spectroscopy mounted on \textit{James Webb Space Telescope} (JWST) in the Abell~2744 and SMACS~J0723-3732 fields. 

This paper is organized as follows. 
We briefly describe the observations, data, and sample selection in Section~\ref{sec:data}.
In Section~\ref{sec:results} we measure and derive the physical properties of our sample and we report the results of MZR for dwarf galaxies at $z=2-3$.
We investigate the physical interpretation of the shape and evolution of MZR and discuss the existence of fundamental metallicity relation in Section~\ref{sec:discuss}.
Throughout this paper, we use AB magnitudes \citep[e.g.,][]{Oke:1983}.
We adopt a flat $\mathrm{\Lambda CDM}$ cosmology with $\Omega_{\mathrm{m}}=0.3$ and $H_{0}=70 \mathrm{~km}\mathrm{~s}^{-1} \mathrm{~Mpc}^{-1}$ and a \citet{Chabrier2003} Initial Mass Function (IMF).

\section{Data and Sample} \label{sec:data}

\begin{figure*}[t]
\centering
\figurenum{1(a)}
\includegraphics[width=\linewidth]{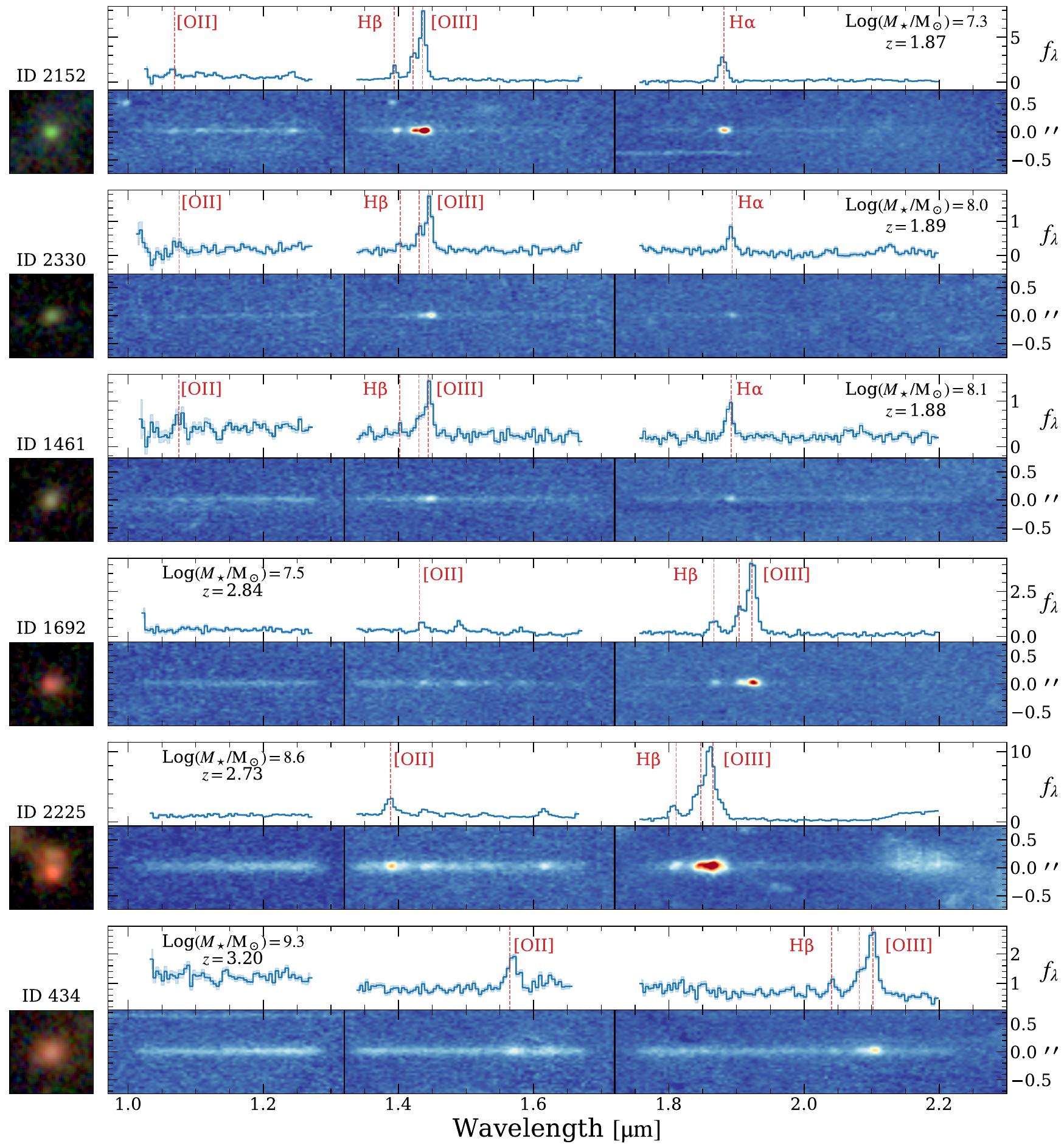}
\caption{The $1.5\arcsec \times1.5\arcsec$ RGB (red for JWST/NIRISS F200W, green for F150W, and blue for F115W) cutouts, extracted 1D spectra, and corresponding 2D spectra for six galaxies in our sample. Their IDs, redshifts, and stellar masses are labeled.
The flux density of 1D spectra, $f_\lambda$, is in unit of $10^{-19}\mathrm{\,erg\,s^{-1}\,cm^{-2}\,\AA^{-1}}$.}
\label{fig:emitter_rgb}
\end{figure*}

\begin{figure*}
\centering
\figurenum{1(b)}
\includegraphics[width=\linewidth]{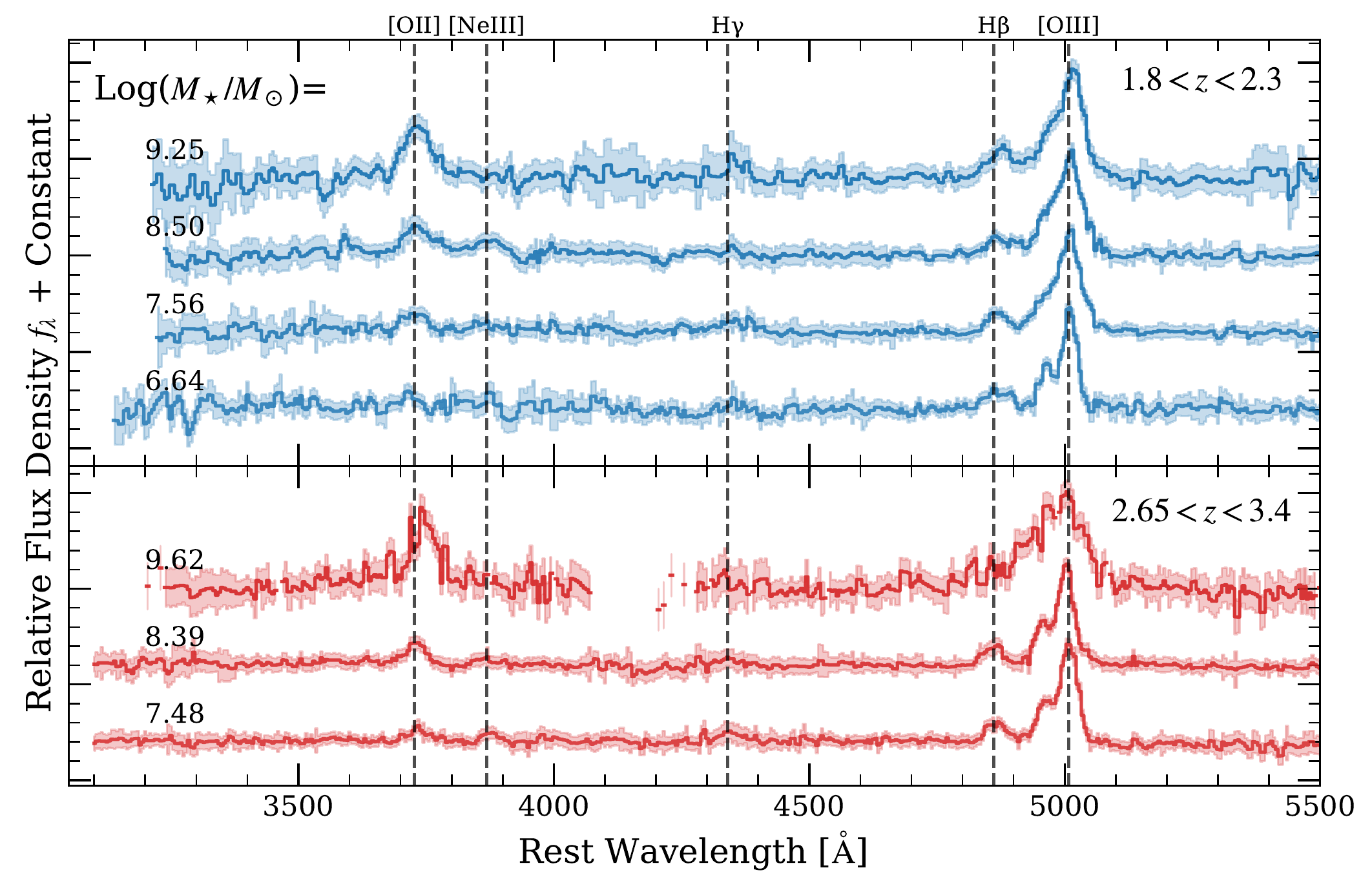}
\caption{One-dimensional spectral stacks binned by redshift and stellar mass, with each individual continuum-subtracted spectrum normalized by a de-reddening flux of
[\ion{O}{3}] (see Section~\ref{sec:Z} for more details). The solid lines show the weighted average stacked spectra, and the shaded regions indicate the corresponding 1$\sigma$ uncertainties. The low S/N wavelength ranges located at the edge of filter bands are masked. Emission lines are labeled as black dashed lines. The [\ion{O}{3}]/[\ion{O}{2}] line ratio evolves prominently with the stellar mass, demonstrating that galaxies with lower stellar mass tend to be more metal-poor. 
}\label{fig:stack_spec}
\end{figure*}

We create our galaxy sample in two cluster fields, Abell~2744 (A2744) field and SMACS~J0723-3732 (SMACS~0723).
These fields are extensively covered by JWST/NIRISS grism spectroscopy and imaging data from HST and JWST.
The wavelength range (0.9 - 2.3~$\mu$m) of the JWST/NIRISS spectroscopy of these galaxies covers the metallicity diagnostics emission lines, including [\ion{O}{2}], 
[\ion{O}{3}], H$\beta$ at $z=2-3$.
The high signal-to-noise (S/N) of JWST/NIRISS grism spectroscopy plus gravitational lensing allows us to detect the emission lines of galaxies with extraordinarily low stellar masses ($\approx$ $10^{6.5}\ \mathrm{M}_\odot$ at $z=2-3$). 
The complete coverage of the optical and IR imaging in SMACS~0723 \citep[e.g.,][]{Pontoppidan2022,Coe2019} and A2744 fields \citep[e.g.,][]{Treu2022,Lotz2017} also allows
us to have an excellent estimate of the stellar mass down to $10^6\ \mathrm{M}_\odot$ at $z=2-3$, allowing for our measure of the MZR using a sample of dwarf galaxies at $z=2-3$.

\subsection{JWST/NIRISS Grism Spectroscopic Observations}\label{sec:grismData}

NIRISS observations of the A2744 field were taken by the GLASS JWST ERS program \citep[ERS-1324:][]{Treu2022}.
The integration time is $2835\mathrm{~s}$ ($\simeq 47\mathrm{\,minutes})$ for pre-imaging and $2\times5196.5\mathrm{~s} ~(\simeq 2.9\mathrm{\,hr})$ for GR150 row and column (R+C) grism with three filters (F115W, F150W, F200W). This observation setup provides almost continuous wavelength coverage from $1$ to $2.2\mathrm{~\mu m}$ in the A2744 field.

SMACS~0723 field was targeted by the JWST early-release observation program \citep[ERO 2736:][]{Pontoppidan2022}. The JWST/NIRISS grism spectroscopy was obtained in F115W (1--1.3\,\micron) and F200W (1.7--2.2\,\micron) filters.
The exposure time of the NIRISS grism observations of SMACS~0723 is $2\times 2834.5\mathrm{~s} ~(\simeq 1.6 \mathrm{~hr})$ for each {filter}, respectively. 

The JWST data were reduced and calibrated following the methodology of \citet{Wu2022}.
The data were reduced using the standard JWST pipeline\footnote{\url{https://github.com/spacetelescope/jwst}} v1.9.4 with calibration reference files ``\texttt{jwst\_1041.pmap}''.
First, we modeled the 1/f noise (see \citealt{Schlawin20}) and removed it using the \texttt{tshirt/roeba} code\footnote{\href{https://github.com/eas342/tshirt}{https://github.com/eas342/tshirt}}.
Then, we identified ``snowball'' artifacts from cosmic rays \citep{Rigby22} and masked them.
We then registered the world coordinate system of mosaicked images using the Pan-STARRS DR1 catalog \citep{panstarrsdr1} \citep[Gaia DR2 catalog,][]{GaiaCollaboration2018} for A2744 (SMACS~0723) field.
Lastly, the pixel scale of final mosaicked images was resampled to 0.03\arcsec\ with \texttt{pixfrac}\,$=$\,0.8. We reduced NIRISS grism data using {the} Grism Redshift \& Line {Analysis tool} (\texttt{Grizli}\footnote{\url{https://github.com/gbrammer/grizli}}; \citealt{Brammer2022}).
The 2D grism spectra were drizzled with a pixel scale of 0.065\arcsec.

The JWST/NIRISS grism observations cover the metallicity-senstive lines [\ion{O}{2}]$~\lambda\lambda 3727,3729$, H$\beta$, and [\ion{O}{3}]$~\lambda\lambda4959, 5007$, allowing for our gas-phase oxygen abundance estimate.
The spectral resolution ($R\equiv\lambda/\Delta\lambda$) of JWST/NIRISS observations is $\approx$150, enabling the separation of the [\ion{O}{3}]$~\lambda\lambda4959, 5007$ doublet.

We used \texttt{Grizli} to perform target detection, contamination modeling, and spectrum extraction.
First, we constructed pre-imaging mosaics, which were used to generate catalogs of sources and associated segmentation maps. 
Then, we built contamination models for all sources by forward modeling, which eliminates both zeroth and higher-order spectra from nearby sources. Finally, we extracted contamination-removed spectra for detected sources.

We determined galaxy redshifts by spectral template synthesis (SSP) fitting.
From the extracted sources, we generated a catalog of galaxies with robust redshift solutions, which required only one significant solution ($\log(p(z)/\chi^2)>1$) for redshift template fitting.
Specifically, at $z\simeq2-3$, we expect multiple strong emission lines within the NIRISS wavelength coverage, e.g., [\ion{O}{2}], H$\beta$, [\ion{O}{3}], and H$\alpha$.
The redshifts can be well fitted by the resolved [\ion{O}{3}] doublets and the secondary lines (e.g., [\ion{O}{2}], H$\beta$, and H$\alpha$) allow us to perform a more robust redshift fitting.

\subsection{HST and JWST Photometric Imaging}\label{sec:photometry}

Because A2744 cluster is one of the Hubble Space Telescope Frontier Fields (HFF) program\footnote{\url{https://archive.stsci.edu/prepds/frontier/}} \citep[e.g.,][]{Lotz2017} and the SMAC~0723 field is part of the Reionization Lensing Cluster Survey: RELICS\footnote{\url{https://archive.stsci.edu/prepds/relics/}} \citep[][]{Coe2019}, each cluster has extensive archival photometric imaging across a range of wavelengths. 
Specifically, these include HST/ACS in F435W, F606W, F814W filters, and WFC3/IR in F105W, F125W, F140W, F160W filters. In this study, we use the high-level science products released by Space Telescope Science Institute (STScI) with a pixel size of 30 mas.

Both A2744 and SMACS~0723 have grism pre-imaging from JWST/NIRISS as references to extract the grism spectra (see Section~\ref{sec:grismData}).
Additionally, SMACS~0723 field has deep JWST/Near Infrared Camera 
(NIRCam) imaging data in the F090W, F150W, F200W, F277W, F356W, and F444W bands.

For HST imaging in the A2744 field, we used photometric results from the HFF-DeepSpace multiple wavelength catalogs \citep[e.g.,][]{Shipley2018, Nedkova2021}.
For HST imaging in the SMACS~0723 field and JWST imaging data in both fields, we use \texttt{SExtractor} \citep[][]{Bertin1996} to measure Kron-like \texttt{AUTO} photometry of the galaxies on our custom calibrated JWST images and public HST data product.
We performed individual measurements and subsequently crossmatched them to obtain a photometric catalog using the detection of NIRISS F200W as the reference.
Finally, we built the combined catalog with photometry in F435W, F606W, F814W, F105W, F115W, F125W, F140W, F150W, F160W, and F200W bands for both fields.
For the SMACS~0723 field, we also include the F090W, F277W, F356W, and F444W photometry.
A 5\% photometry relative uncertainty floor is set for JWST bands to involve the systematic zero-point error.
Based on this photometric catalog, we perform spectral energy distribution (SED) fitting to derive the physical properties of each galaxy, as explained in Section~\ref{sec:SED}.

\subsection{Sample Selection} \label{sec:sample}
We selected star-forming galaxies based on the JWST/NIRISS wide field slitless spectroscopy.
We used the following criteria to select galaxies from the source catalog as established 
in Section~\ref{sec:grismData}.

\begin{itemize}
    \item With the redshifts determined by NIRISS grism, we select galaxies at $1.8 < z < 2.3$ and $2.65 < z < 3.4$.  For the SMACS~0723 field, the absence of grism observations in the F150W band limited us to only select galaxies at $2.65 < z < 3.4$. These redshift ranges ensure that NIRISS spectra completely cover the [\ion{O}{2}], H$\beta$, and [\ion{O}{3}] emission lines in the A2744 field, and H$\beta$ and [\ion{O}{3}] emission lines in the SMACS~0723 field. Our redshift selection criteria also ensure these emission lines are located in the high response wavelength interval ($>90\%$ response at filter center), which avoids the impact of large flux calibration uncertainty. Based on this redshift criterion, we selected 76 galaxies to be the primary sample to apply the following criteria.

    \item The signal-to-noise ratio of the [\ion{O}{3}] line should be greater than ten. This criterion was established by empirical inspection. We did not require the signal-to-noise of secondary lines, which could lead to the Eddington bias\citep{Eddington1913}. The galaxies that have secondary line detection (e.g., H$\beta$ or [\ion{O}{2}]) tended to have higher metallicity \citep[e.g.,][]{Henry2021}. To minimize this selection effect, only [\ion{O}{3}] detection significance was required instead. 72 galaxies satisfied this criterion.

    \item To remove AGN contamination, we excluded galaxies that fall in the ``AGN regime" of the mass–excitation (MEx) diagram ([\ion{O}{3}]$~\lambda5007$/H$\beta$ versus $M_\star$) defined by \citet{Juneau2011} and modified by \citet{Coil2015}. In all galaxies, we found H$\beta$ to be at least marginally detected ($> 1.5\sigma$), so there are very few sources with ambiguous [\ion{O}{3}]/H$\beta$ on the MEx diagram. Eight galaxies were classified in the AGN-like category and were removed from our sample. Because we detected no other AGN diagnostic lines (e.g., [\ion{S}{2}] and [\ion{N}{2}]), the MEx diagnostic was the best method available to remove AGN candidates from our sample.

    \item We performed visual inspections on each individual galaxy and removed heavily contaminated sources. Due to the challenges of extracting the spectrum of Multi-images and arclets, we removed these sources following \citet[][]{Mahler2018, Bergamini2022}. 
\end{itemize}

Ultimately, we selected 51 galaxies, where 44 are in A2744 and seven in SMACS~0723.
We show six examples of the JWST/NIRISS color composite stamps and spectra in Figure~\ref{fig:emitter_rgb}.
Among all galaxies in the sample, 27 galaxies are at redshift $1.8<z<2.3$ ($z_\mathrm{med} = 2.01$), and 24 galaxies are at redshift $2.65<z<3.4$ ($z_\mathrm{med} = 2.97$).
The redshifts of 42 galaxies are spectroscopically determined by NIRISS observations for the first time, 
while the remaining nine galaxies have been previously confirmed by HST/WFC3 or VLT/MUSE \citep[e.g.,][]{Wang2020, Mahler2018, Boyett2022}.
Among the nine galaxies, the NIRISS redshifts of two galaxies (ID 1692 and ID 2225) are inconsistent with those in previous studies.
For these two galaxies, we re-checked their spectra (shown in Figure\ref{fig:emitter_rgb}) and confirm the reliability of the NIRISS redshifts based on the appearance of multiple emission lines with S/N $>$ 3.

\section{Analysis and Results} \label{sec:results}

\begin{figure*}[t]
    \centering
    \figurenum{2}
    \includegraphics[width=0.9\linewidth]{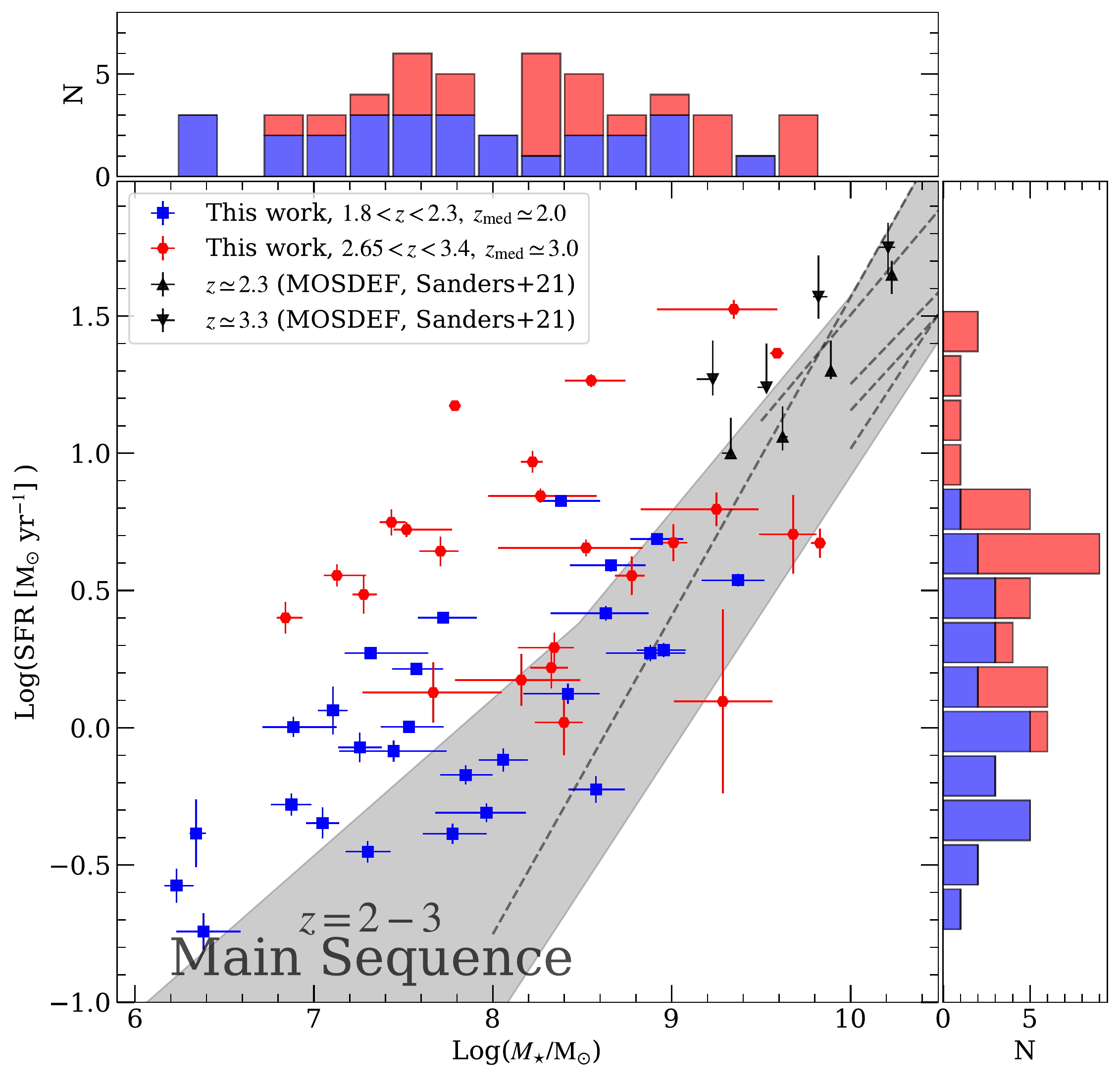}
    \caption{The distribution of galaxies in our sample on the SFR–$M_\star$ plane. The red squares show the $z \simeq 2$ sample and the red hexagons show the $z \simeq 3$ sample. The MOSDEF stacked results are denoted in black triangles.  The gray shaded regions shows the (extrapolated) galaxy main sequence at $z=2-3$ based on the results from \citet{Speagle2014}, \citet{Schreiber2015}, \citet{Santini2017}, \citet{Pearson2018}, and \citet{Leja2022}. The origin results from these works are denoted as gray dashed lines. The top and right panels show the stellar mass and SFR distributions respectively, with the $z\simeq2$ sample in red and $z\simeq3$ sample in blue.  
    }\label{fig:SFRMS}
\end{figure*}

Based on the spectral and photometric data, we measure and derive the galaxy properties from the emission line intensities and the SED fitting using multiple-band photometry. In this section, we present the methodology for the measurement and derivation of galaxy properties and show the results of MZR.

\subsection{Emission-line Flux}\label{sec:lineflux}
To measure the intensity of galaxy emission lines, we use the Gaussian profile models to fit the lines in continuum-subtracted grism spectra.
The line wavelengths are fixed by the galaxy's redshift, which is determined by the SSP fitting as presented in Sec~\ref{sec:grismData}.
For the H$\alpha$ and [\ion{O}{2}] lines, we fit the spectra using a single 1D Gaussian function.
For the H$\beta$+[\ion{O}{3}]$\lambda\lambda4959,5007$ lines, we fit the spectra using three Gaussian functions.
We fix the line ratio of the [\ion{O}{3}] doublet to [\ion{O}{3}]$\lambda4959$/[\ion{O}{3}]~$\lambda5007=1/3$ and require the line width of [\ion{O}{3}] doublet to be equal.
We further verify the accuracy of our derived flux by comparing our measurements to the modeled flux solution fitted by \texttt{Grizli} and find the two to be consistent.
A thorough summary of the best-fit observed emission line fluxes can be found in Appendix~\ref{sec:property}.

In the early era of JWST observations, data calibration cannot be completely accurate \citep[e.g.,][]{Rigby2022}. To minimize calibration issues, we employed reference calibration files that have been corrected using space-based observations. 
To quantify the effect of any remaining systematic errors on our data, we compared the NIRISS-extracted spectra with previous HST/WFC3 grism observations. We found that the estimated relative flux calibration error between different filter bands was limited to $\lesssim5\%$, which is consistent with values reported in previous literature \citep[e.g.,][]{Wang2022}.

\subsection{Stellar Mass and Star Formation Rate}\label{sec:SED}

Galaxy stellar masses are derived by SED fitting to the photometry shown in Sec~\ref{sec:photometry}.
Note that both A2744 and SMACS~0723 clusters significantly magnify the background sources, including galaxies in this work due to the gravitational lensing.
In order to calculate the lensing magnification of each galaxy at the appropriate redshift, we adopt lensing models from \citet{Mahler2018} and \citet{Golubchik2022}, respectively (see Appendix~\ref{sec:property}).
To do our analysis, we first correct the magnitudes in all bands by the lensing magnification factor. Note that we do not introduce the systematic error of lensing models.
To simultaneously measure the stellar mass ($M_{\star}$), the stellar dust attenuation ($\rm E(B-V)_{star}$), and the best-fit model of the stellar {continuum} with emission lines, we fit the SED using the Bayesian code BEAGLE \citep[][]{Chevallard2016}. 
For our SED fitting, we fix the galaxy redshift, assume a constant star formation history, and model the dust attenuation curve following \citet[][]{Calzetti2000}.
We set the dust attenuation in the $\mathrm{V}$ band $A_\mathrm{V}$ to vary from $0-8$ and use the \citet{Chabrier2003} initial mass function (IMF) with an upper limit of $300~\mathrm{M}_{\odot}$. 
The metallicity of the interstellar medium (ISM) is assumed to be the same as that of 
the stellar populations.
The nebular line and continuum emission are included in our fitting based on the \texttt{CLOUDY} {photoionization code} \citep[][]{Ferland2017}.
To model the metallicity, we use a flat prior with values ranging from  $-2 <\log(Z / \mathrm{Z}_{\odot})<0.4$.
For our model, we bound the ionization parameter in log-space between $-4<\log U<-1$.

The derivation of star formation rate (SFR) and gaseous metallicity is based on the attenuation-corrected emission line intensities.
We correct the reddening effect of dust attenuation for all measured nebular lines using the $\rm E(B-V)_{star}$ which is derived from the SED fitting, assuming the dust reddening for the nebular and stellar components are equal. Indeed, several works \citep[e.g.,][]{Shivaei2020} pointed out that $\rm E(B-V)_{star}$ could equal to $\sim 2\rm E(B-V)_{gas}$ for galaxies at $z=2$, we test using this assumption and found out that the SFR could be increased by 0.1 dex. Furthermore, the relative changing line ratios (e.g., $\rm O_{32}$) are at an average level of 0.03 dex, which will lead to consistent conclusions in this paper. Therefore, in the rest of the paper, we adopt the assumption of $\rm E(B-V)_{star}=\rm E(B-V)_{gas}$.

Note that the nebular dust attenuation can be also derived from H$\alpha$/H$\beta$ Balmer decrement (e.g., H$\alpha$/H$\beta$ or H$\beta$/H$\gamma$) for a few galaxies in our sample. But half of our galaxy sample does not have H$\alpha$ coverage. The absence of H$\alpha$ line coverage and lack of significant H$\gamma$ line detection makes it impossible to measure dust attenuation via the Balmer decrement. To maintain consistency among our entire sample, we adopt the dust attenuation from the SED fitting to perform the de-reddening correction for emission lines.

We measure the SFR using the attenuation-corrected H$\alpha$ luminosity.
For galaxies without H$\alpha$ coverage, we derive the H$\alpha$ luminosity from the attenuation-corrected H$\beta$ flux by adopting a Balmer decrement of H$\alpha$/H$\beta$=2.86.
In order to estimate the SFR we assume the \citet{Kennicutt1998} SFR calibration and the \citet{Chabrier2003} initial mass function.
The SFR is modeled as
\begin{equation}
    \mathrm{SFR} = 4.6\times10^{-42}\frac{L_\mathrm{H\alpha}}{\rm erg~s^{-1}} [\mathrm{M}_\odot~\mathrm{yr}^{-1}],
\end{equation}
where $L_\mathrm{H\alpha}$ {is} the H$\alpha$ luminosity.
Throughout this paper, SFR corresponds to the $\rm SFR_{H\alpha}$, which is derived from the attenuation-corrected H$\alpha$ luminosity.
And for those galaxies without H$\alpha$ coverage, we get H$\alpha$ luminosity using the attenuation-corrected H$\beta$ assuming a Balmer decrement of H$\alpha$/H$\beta$=2.86.

Within our sample, it is worth mentioning that the H$\alpha$ line is blended with [\ion{N}{2}] lines due to the low spectral resolution, which biases our estimate of H$\alpha$ towards a higher value. 
However, the contribution of [\ion{N}{2}] decreases with metallicity and typically the [\ion{N}{2}]/H$\alpha$ is less than 0.07 for star-forming galaxies with the stellar mass of $M_\star<10^{9}\mathrm{M}_\odot$ at $z=2-3$ \citep[e.g.,][]{Bian2018}.
Therefore, the [\ion{N}{2}] contamination should be negligible to our SFR measurement, especially when compared with the uncertainty of dust attenuation correction.
Besides, Balmer line stellar absorption can bias the intrinsic line luminosity to a lower value.
Nevertheless, for the young star-forming galaxies in our sample, the Balmer line stellar absorption is comparatively weak and marginalized.

Using the measurements from SED fitting and SFR calibration, we plot Log(SFR) as a function of the Log(M$_{\star}$), in Figure~\ref{fig:SFRMS}.
We find that the relation between SFR and $M_\star$ displays a positive correlation across the complete range of stellar masses and redshift.
We also find this slope agrees with the more massive galaxies measured in the MOSDEF sample \citet{Sanders2021} and the extrapolated galaxy star-forming main sequence.
Despite a large degree of scatter, the SFR at fixed $M_\star$
increases
with the redshift (Figure~\ref{fig:SFRMS}). 
However, when compared with previous studies of star-forming main sequence galaxies at redshifts of $2<z<3$, our sample of galaxies has a slightly higher mean SFR.

\begin{deluxetable*}{ccccccc}
\tablenum{1}
\tablecaption{Properties of stacked spectra in bins of $M_\star$ for the $z\simeq2.0$ and $z\simeq3.0$ samples.\label{tab:stack}}
\tablewidth{0pt}
\tablehead{
\colhead{log$(M_\star/\mathrm{M}_\odot)$} & \colhead{N} & \colhead{($\rm 12+ log(O/H)$)$_\mathrm{stack}$} & \colhead{($\rm 12+ log(O/H)$)$_{\rm med}$} & \colhead{$\rm O_{32}$} & \colhead{$\rm O_{3}$} & \colhead{SFR [$\mathrm{M_\odot~yr^{-1}}$]}
}
\decimalcolnumbers
\startdata
\multicolumn{7}{c}{\large $1.8<z_\mathrm{grism}<2.3$}\\
$6.64\pm{0.12}$ &  8 & $8.03\pm{0.05}$ & $8.00\pm{0.10}$ & $0.86\pm{0.08}$ & $0.59\pm{0.04}$ & $0.61\pm{0.12}$ \\
$7.56\pm{0.09}$ &  9 & $8.11\pm{0.03}$ & $8.05\pm{0.06}$ & $0.73\pm{0.06}$ & $0.79\pm{0.04}$ & $1.09\pm{0.25}$ \\
$8.50\pm{0.09}$ &  6 & $8.23\pm{0.03}$ & $8.26\pm{0.06}$ & $0.52\pm{0.05}$ & $0.74\pm{0.04}$ & $2.61\pm{0.95}$ \\
$9.25\pm{0.20}$ &  4 & $8.35\pm{0.02}$ & $8.38\pm{0.07}$ & $0.32\pm{0.03}$ & $0.55\pm{0.03}$ & $2.83\pm{0.57}$ \\
\hline
\multicolumn{7}{c}{\large $2.65<z_\mathrm{grism}<3.4$}\\
$7.48\pm{0.09}$ &  7 & $8.00\pm{0.04}$ & $7.98\pm{0.08}$ & $0.91\pm{0.06}$ & $0.75\pm{0.03}$ &  $5.02\pm{1.77}$ \\
$8.39\pm{0.10}$ &  6 & $8.14\pm{0.03}$ & $8.10\pm{0.05}$ & $0.67\pm{0.04}$ & $0.78\pm{0.03}$ &  $5.60\pm{2.35}$ \\
$9.62\pm{0.06}$ &  4 & $8.34\pm{0.02}$ & $8.40\pm{0.09}$ & $0.34\pm{0.03}$ & $1.01\pm{0.10}$ & $10.64\pm{6.97}$ \\
\enddata
\tablecomments{Column 1 is the median stellar mass for each bin; Column 2 is the number of galaxies in each bin; Columns 3 and 4 are gas phase metallicity measured from the stacked spectra and statistical median of individual galaxies, respectively; Column 5 and 6 are the line ratios O$_{32}=$ log([\ion{O}{3}]$\lambda\lambda4959,5007$/[\ion{O}{2}]$\lambda\lambda3727,3729$) and $\mathrm{O}_3$=log([\ion{O}{3}]$\lambda\lambda4959,5007$/H$\beta$) derived from the stacked spectra; Column 7 is the median star formation rate.}
\end{deluxetable*}

\subsection{Gas-phase Metallicity}\label{sec:Z}

\citet{Bian2018} established empirical relations between the strong metallicity diagnostic line ratios and the direct-$\mathrm{T_e}$ metallicity using a sample of local analogs of high-redshift galaxies.
These calibrations have been demonstrated as one of the most reliable metallicity calibrations at $z>1$ \citep{Sanders2020}.  

We derive the gas-phase oxygen abundance by adopting the \citet{Bian2018} calibration and measuring the strong metallicity diagnostic line ratios O$_{32}=$ log([\ion{O}{3}]$\lambda\lambda4959,5007$/[\ion{O}{2}]$\lambda\lambda3727,3729$) and $\mathrm{O}_3$=log([\ion{O}{3}]$\lambda\lambda4959,5007$/H$\beta$). 
This calibration allows us to minimize the systematic uncertainty of the metallicity measurements when comparing our MZR results with those from \citet{Sanders2021}.
For the galaxies with both [\ion{O}{3}] and [\ion{O}{2}] coverage {(44/51)}, we use the $\mathrm{O_{32}}$ to measure the gas-phase oxygen abundance as follows
\begin{equation}\label{equ:Z-O32}
    12 + \mathrm{log(O/H)} = 8.54 - 0.59 \times \mathrm{O_{32}}.
\end{equation}

As shown above, we use the \citet{Bian2018} calibration, which is suitable in the range of $0.3<\mathrm{O_{32}}<1.2$ to measure the oxygen abundance in our galaxies. For the galaxies (11/48) where $\rm O_{32}$ is out of this range, we extrapolate this relation linearly to estimate their oxygen abundance.
The linearity of the O$_{32}$ calibration allows us to extrapolate with minimal systematic error.
For the same reason, we try not to use the $\rm R_{23}$ and $\rm O_3$ to further constrain the metallicity, which will cause more systematic uncertainties, especially for low-mass and low-metallicity galaxies.

For galaxies without [\ion{O}{2}] coverage (7/51), we calculate the metallicity by using line ratio $\mathrm{O_3} \equiv $ [\ion{O}{3}]$\lambda\lambda$4959,5007/H$\beta$, and use the following equation:

\begin{equation}
\mathrm{O_3}=43.9836-21.6211 x+3.4277 x^2-0.1747 x^3,
\end{equation}
where $x=12+\mathrm{log(O/H)}$. 

For the above $\mathrm{O_3} - Z$ relation, there exist two oxygen abundance solutions for a given $\mathrm{O_3}$ value.
We adopt the {solution} within the range of $7.8<12+\log(\mathrm{O/H})<8.4$, which is consistent with the oxygen abundance range derived from $\mathrm{O_{32}}$ indicator in a similar mass range.
Our derived metallicities {are} presented in Appendix~\ref{sec:property}.

Additionally, we stack each individual spectrum to increase the signal-to-noise and measure the average metallicity in different stellar mass and redshift bins.
We perform the spectrum stacking in two redshift bins ($1.8<z<2.3$ and $2.65<z<3.4$). 
For our stacking analysis, we only use the galaxies in the A2744 field, because those in the SMACS~0723 field do not have [\ion{O}{2}] coverage.
We divide the entire galaxy sample into four (three) stellar mass bins within the redshift range $1.8 < z < 2.3$ ($2.65<z<3.4$). 
Each stellar mass bin contains between four and nine galaxies, which is determined by the bootstrap simulation to minimize the uncertainty of derived metallicities of stacked spectra. Moreover, we ensure uniformity of the median stellar mass of each mass bin when setting the galaxy number in each bin.

For our stacking analysis, we first shift each spectrum into the rest frame according to the grism redshift.
We then apply dust-corrections following the E(B-V) values from our SED fitting and the \citet{Calzetti2000} attenuation curve.
We then subtract off the continuum for each individual spectrum before normalization and stacking. We normalized each spectrum based on their [\ion{O}{3}] flux and resampled it onto a uniform wavelength grid.
Lastly, the spectra in each mass bin are combined by taking the mean of the 3-$\sigma$ clipping of each wavelength grid.
In Figure~\ref{fig:stack_spec}, we present the final stacked spectra in different stellar mass and redshift bins. 
We measure the relative emission line flux in the stacked spectra using the same method as for the individual galaxies.
We derive gas-phase metallicities using the $\mathrm{O}_{32}$ indicator.
The measurement uncertainties are propagated from both the stacking and line flux fitting. 
The median stellar mass, measured line ratios, median SFR, and the derived metallicity of stacked spectra are presented in Table~\ref{tab:stack}.

It's important to note that performing the dust correction either before or after stacking during the stacking procedure could potentially introduce bias to the measured metallicity\citep[e.g.,][]{Henry2021}.
We performed a test by changing the sequence of dust corrections.
Consequently, we find that the difference in derived metallicity 12+log(O/H) is 0.01 - 0.03 typically, which is within $1\sigma$ error of metallicity $\simeq 0.02-0.04$.
Therefore, our results are not sensitive to the dust correction sequence when stacking.
We also measure the dust attenuation using the Balmer decrement on the stacked spectra that are not dust-corrected.

\begin{figure*}[t]
    \figurenum{3}
    \centering
    \includegraphics[width=0.9\linewidth]{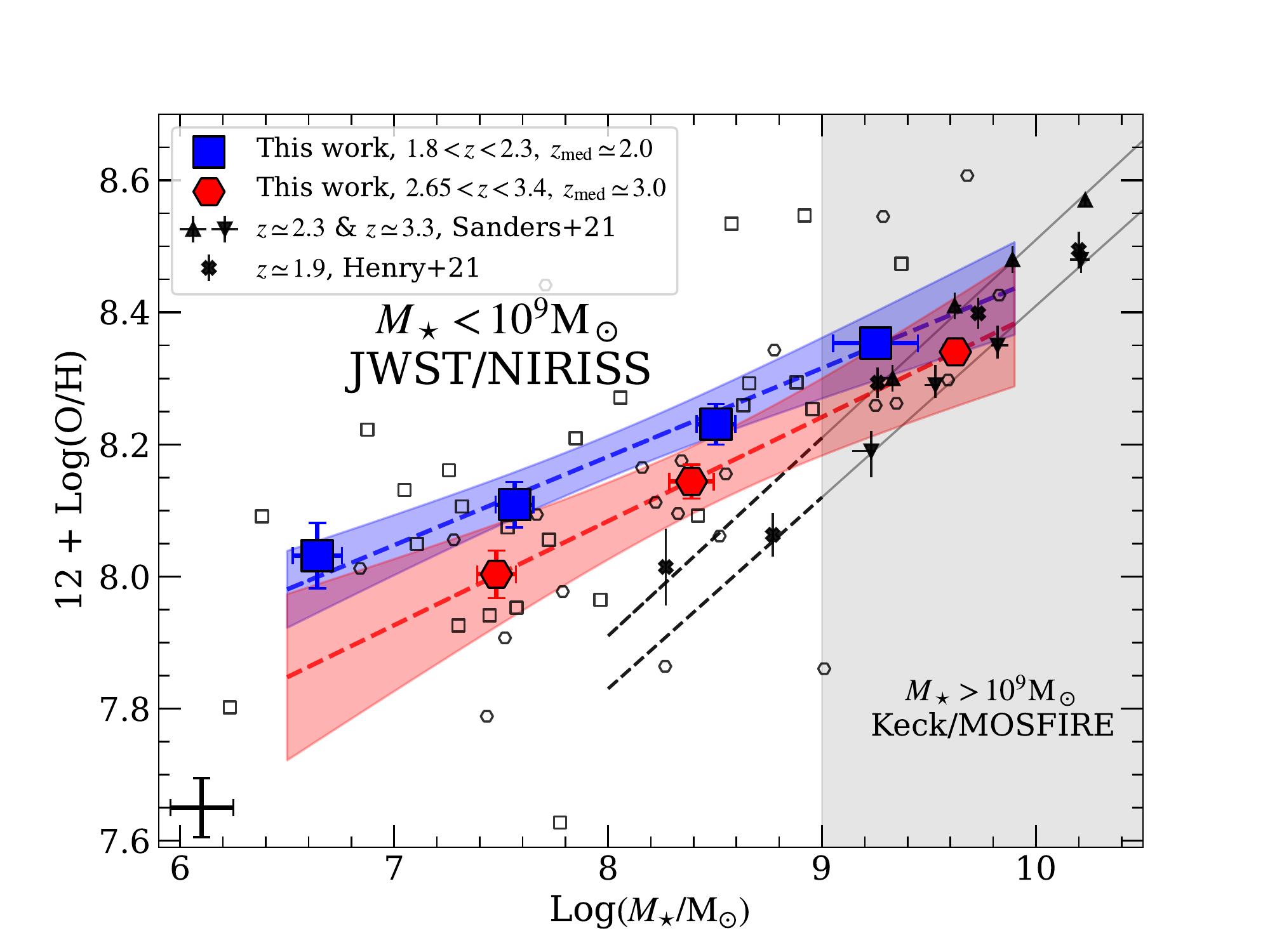}
    \caption{The mass-metallicity relation (MZR) for dwarf galaxies at $z\simeq 2$ (blue) and $z\simeq 3$ (red). The unfilled squares and hexagons denote individual galaxies at $z\simeq2$ and $\simeq3$ respectively. The large blue squares and red hexagons mark the results based on stacked spectra.
    The error bar in the lower left corner displays the median uncertainty of the individual galaxies. The best-fit results are denoted by the colored dashed lines, with their $1\sigma$ uncertainties in the shaded region.The MOSDEF stacked results \citep{Sanders2021} are marked as black triangles and \citet{Henry2021} stacked results are marked as black crosses. The MOSDEF best-fit result and the extrapolation to the low-mass end are in solid and dashed lines.}
    \label{fig:MZR}
\end{figure*}

\begin{figure*}[t]
    \figurenum{4}
    \centering
    \includegraphics[width=0.9\linewidth]{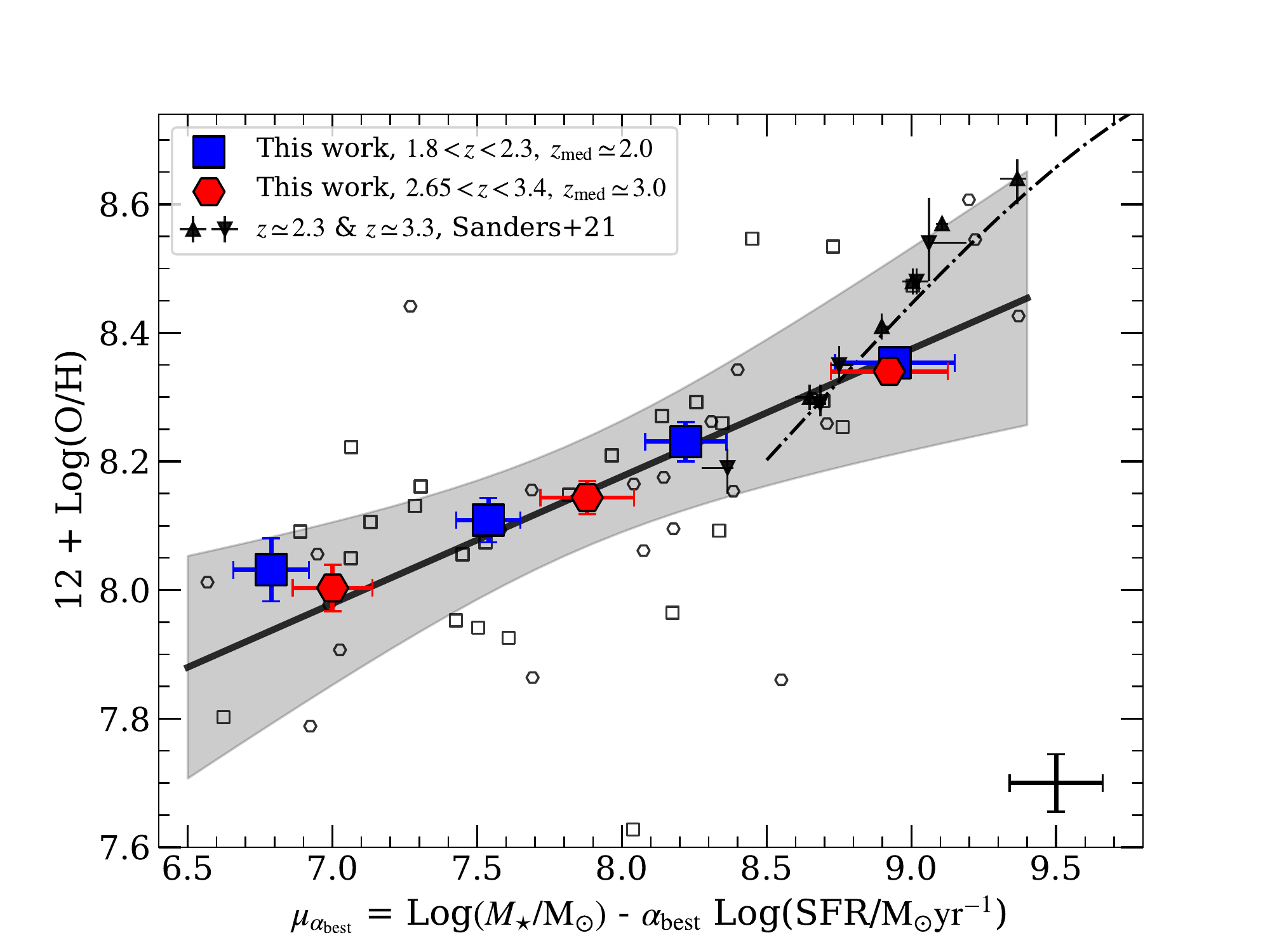}
    \caption{Projection of the fundamental metallicity relation (FMR) as O/H vs. $\mu_{\alpha_\mathrm{best}} = \operatorname{log}(M/\mathrm{M_\odot}) - \alpha_\mathrm{best}\times \operatorname{log}(\mathrm{SFR/M_\odot~yr^{-1}})$. The coefficient $\alpha_\mathrm{best}=0.60$, minimizes the scatter of the O/H for our sample. The solid black line shows the best linear fit from Equation~\ref{equ:FMR} and the grey-shaded region shows the $1\sigma$ uncertainty. The unfilled squares and hexagons represent individual galaxies at $z\simeq2$ and $z\simeq3$ respectively, and the large colored ones show the stacked results. The black triangles denote the MOSDEF stacked results from \citet{Sanders2021} and the black dotted-dashed line marks their best-fit. The error bar in the lower right corner displays the median uncertainty of the individual galaxies.}
    \label{fig:FMR}
\end{figure*}

\subsection{The Mass–Metallicity Relation} \label{sec:MZR}

We find a clear correlation between gas-phase metallicity and stellar mass for our measurements of both individual galaxies and composite spectra.
This is the first-ever effort to derive the MZR down to galaxies of $M_\star = 10^{6.5} \mathrm{M}_\odot$ at $z=2-3$ using grism slitless spectroscopy.
Within both the $z\simeq2$ and $z\simeq3$ sample, we find Spearman correlation coefficients of 0.70 and 0.68, respectively, with the statistical-significant p-value $<10^{-4}$. This means a positive correlation with strong evidence to reject the null hypothesis.

Beyond the strong trends of the MZR within each sample, we also look for signs of redshift evolution.
Using the stacked spectra for the two samples, we find that the MZRs at $z \simeq 2.0$ and $z \simeq 3.0$ show a monotonic evolution 
toward lower metallicity with increasing redshift. 
As seen in Figure~\ref{fig:MZR}, the scale of this evolution is similar to that of MOSDEF 
from $z \simeq 2.3$ to $z \simeq 3.3$.

To quantify the MZR and derive slope and normalization, we fit it using a power law of the form:
\begin{equation}\label{equ:FMR}
    \mathrm{12+log(O/H)} = \beta \times \mathrm{log}\left(\frac{M_\star}{10^{8}~\mathrm{M}_\odot}\right) + Z8,
\end{equation}
where $Z8$ is the metallicity at the stellar mass of 
$10^{8}~\mathrm{M}_\odot$.
Using all galaxies in the sample, we get the best-fit relations with $1 \sigma $ uncertainties are: 
\begin{equation}
\begin{aligned}
    \mathrm{12+log(O/H)} &=\\
    (0.14\pm 0.04) \times &\mathrm{log}\left(\frac{M_\star}{10^{8}~\mathrm{M}_\odot}\right) + (8.14\pm 0.03).
\end{aligned}
\end{equation}
We also fit the stacked spectra in the $z \simeq 2.0$ and $z \simeq 3.0$.
The best-fit parameters with $1\sigma$ uncertainties are as follows:
for $z\simeq2.0$, $\beta_{z\simeq2} = 0.13\pm 0.04$, $Z8_{z\simeq2} = 8.18\pm 0.02$;
and for $z\simeq3.0$, $\beta_{z\simeq3} = 0.16\pm 0.03$, $Z8_{z\simeq3} = 8.08\pm 0.03$. 
These two best-fit relations and the M-Z values of each individual galaxy are shown in Figure~\ref{fig:MZR}.

Note that the slope of best-fit MZR for all stacked spectroscopy is $0.14\pm 0.04$, 
consistent with both $z\simeq2$ and $z\simeq3$ best-fit slopes, implying that there is no significant {evolution in the} slope between $z=2-3$.
In contrast to the lack of evolution in the slope, we do find moderate evolution in the MZR normalization with redshift between $z=2-3$.
Specifically, we find that the difference in the normalization at $z\simeq 2$ and 
$z\simeq3$ is $\mathrm{\Delta log(O/H)}/\mathrm{\Delta}z = -0.09\pm0.06$. 
Beyond redshift evolution, we also probe for differences in the MZR as a function of stellar mass.
We find that at the low-mass end, we measure a much shallower slope than at the high-mass end mass where $\beta \simeq 0.30$ \citep{Sanders2021}.
This suggests a transition in the slope of the MZR at $M_\star \simeq 10^9 \mathrm{M}_\odot$.

Although our sample is not large enough to allow a statistically robust study of the scatter of the MZR at the low-mass end, we find a tentative increase of the MZR scatter at the stellar mass less than $10^8 ~\mathrm{M}_\odot$. 
By defining the intrinsic scatter as $\sigma_\mathrm{ins} = \sqrt{\sigma_\mathrm{obs}^2 - \sigma_\mathrm{meas}^2}$, where $\sigma_\mathrm{obs}$ is the observation scatter and $\sigma_\mathrm{meas}$ is the measurement uncertainties, we measure the intrinsic scatter to be $0.16-0.18$ dex at $M_\star = 10^8 - 10^9 \mathrm{M}_\odot$.
However, we see this value increase to $\approx 0.23$ dex at $M_\star = 10^7 \mathrm{M}_\odot$ bins implying a tentative evolution in the scatter.

\section{Discussion} \label{sec:discuss}

Using imaging from HST/ACS, HST/WFC3, JWST/NIRISS, and JWST/NIRCAM and grism spectroscopy from JWST/NIRISS, 
we present a study of the MZR in high-$z$ ($z= 2-3$), low-mass galaxies ($10^{6.5}~\mathrm{M}_{\odot} < M_\star < 10^{9.5}\mathrm{M}_\odot$), which probes approximately two orders of magnitude lower than previous studies.
The best-fit slope of the MZR in dwarf galaxies at $z=2-3$ is $0.14\pm0.04$, shallower than that of more massive galaxies ($M_\star > 10^9\ \mathrm{M}_\odot$) at $z\approx2-3$. 
Additionally, we measure moderate evolution in the MZR between the $z\simeq2$ and $z\simeq3$.
The primary difference between our two redshift samples is the normalization, which differs by $\mathrm{\Delta log(O/H)}/\mathrm{\Delta}z = -0.09\pm0.06$.
In the following discussion, we provide a detailed physical interpretation of these results.

\subsection{The MZR slope at low-mass end}

Deep JWST/NIRISS spectroscopy of a sample of dwarf galaxies allows us, for the first time, 
to make a robust measurement of the slope of MZR at $z=2-3$ using low mass galaxies ($M_\star\approx 10^{6.5-9.5}\ \mathrm{M}_\odot$).
At $z= 2 - 3,$ we find a shallower slope in MZR at the low-mass end, in contrast to \citet{Sanders2021}, which examines a sample of galaxies with $M_\star = 10^9 - 10^{11} \mathrm{M}_\odot$. This difference indicates a transition in MZR slope around $M_\star \simeq 10^9 \mathrm{M}_\odot$. 

The differences in our measurements of the MZR slope may be indicative of differences in the feedback mechanisms that regulate it \citep[e.g.,][]{2022ApJ...926...70W, Strom2022}.
Typically, stellar feedback drives the gas and metal out of galaxies, which will result in differences in the gas-phase metallicity of the ISM.
The feedback of stellar winds can be described by the wind mass loading factor ($\eta \equiv \dot{M}_\mathrm{out}/\mathrm{SFR}$), which presents the fraction of gas ejected in winds.
The wind mass loading factor $\eta$ is related to the wind velocity ($v_w$), with $\eta \propto v_w^{-2}$ for the energy-driven wind model, $\eta \propto v_w^{-1}$ for the momentum-driven wind model, and $\eta \equiv \mathrm{const}$ for the constant wind model. 
We argue that the shallower MZR slope at the low-mass end is caused by that galaxies with different masses could be dominated by different feedback mechanisms. 

For example, similar to our analysis, \citet{Torrey2019} used IllustrisTNG simulations and similarly found that the MZR has a shallower slope at the low-mass end.
They argue that their results are indicative that feedback in low-mass galaxies may be dominated by constant wind ($\eta = \rm constant$), while feedback in high-mass galaxies may be dominated by energy-driven wind ($\eta \propto v_w ^2$).
Similarly, \citet{Pillepich2018} presents the necessity to introduce a constant wind model to match the low mass end of the galaxy stellar mass function.
Specifically, the constant wind velocity reduces the wind mass loading factor.
The larger-than-expected wind velocity and lower mass loading factor further result in longer wind recycling times and less direct metal ejection, which manifests as somewhat higher metallicities in the ISM and shallower MZR slope for low-mass galaxies, in agreement with our results. 

Our best fit MZR slope is $\beta = 0.14\pm0.04$, which is consistent with the prediction of the momentum-driven wind model \citep[e.g.,][]{Finlator2008, Guo2016}.
Such a slope suggests that momentum-driven wind is the dominant feedback in dwarf galaxies, in contrast to the constant wind models in \citet{Torrey2019}.
One possible interpretation is that the measured MZR slope in our sample is the average effect of a transition between the energy-driven wind model and constant wind models (the slope of the MZR in momentum-driven wind models is between the other two models).
In this interpretation, the momentum-driven wind model ($\eta \propto v_w^{-1}$) acts as an intermediate state between the energy-driven wind ($\eta \propto v_w^{-2}$) and constant wind ($\eta \equiv \mathrm{const}$).
Thus, a more general power-law wind model could be written as $\eta \propto v_w^{-\alpha_\mathrm{fb}}$ where $\alpha_\mathrm{fb}$ is the power index parameter of the feedback.
Our results indicate that there could exist a continuous transition of wind model power index from $\alpha_\mathrm{fb} = 2$ to $\alpha_\mathrm{fb} = 0$ in low-mass galaxies 
with a transition stellar mass of $M_\star \simeq 10^{9}~\mathrm{M}_\odot$. 

The slope of the MZR at the low-mass end remains not well constrained and is under debate. For example, \citet{Henry2021} found that the MZR slope for galaxies with stellar masses $M_\star>10^8~M_\odot$ at $z\simeq1.9$ could remain steep, similar to that of higher-mass galaxies.
However, the low-mass MZR slope we find is shallower.
The discrepancy with \citet{Henry2021} could arise from several factors, including the different metallicity calibration methods used.
As \citet{Sanders2020} suggested, we employ strong line calibration from \citet{Bian2018} and use $\rm O_{32}$ as the metallicity indicator.
In contrast, \citet{Henry2021} derived metallicities based on \citet{Curti2017} using joint fitting of multiple line ratios.
The choice of strong lines and calibration method could lead to a systematic discrepancy in MZR slope and normalization.
To better compare our results with \citet{Henry2021}, we recalculated the metallicities from their stacked measurements (overlaid in Figure~\ref{fig:MZR}), using our method based on the \citet{Bian2018} calibration.
We find that the stacked results at $M_\star < 10^9 \mathrm{~M_\odot}$ show a tendency to follow the shallower-sloped MZR compared to the higher-mass points.
To better constrain the validity of the slope transition hypothesis, we require further observations of the low-mass MZR and a more robust comparison to simulations.

\subsection{The Evolution of MZR from z=3 to z=2}

For fixed stellar mass $M_\star$, we find that the MZR normalization moderately evolves 
from lower O/H to higher O/H from $z=3$ to $z=2$. 
Moreover, we also aim to determine any evolution with stellar mass by comparing with \citet{Sanders2021}.
For the high-mass end ($M_\star \simeq 10^{10} \mathrm{~M_\odot}$), the normalization evolution is $\mathrm{\Delta log(O/H)}/\mathrm{\Delta}z = -0.11\pm0.02$, while for the result in this work, we found $\mathrm{\Delta log(O/H)}/\mathrm{\Delta}z = -0.09\pm0.06$ ($M_\star \simeq 10^8 \mathrm{~M_\odot}$).
It concludes that the evolution of the MZR is independent of galaxy stellar mass, 
as indicated by an unchanged MZR shape from $z=3$ to $z=2$ in a wide stellar 
mass range of $10^6 - 10^{10}~\mathrm{M}_\odot$.

The evolution of MZR is the consequence of a complicated interplay between star formation efficiency, galactic feedback, gas fraction, gas accretion, and recycling \citep[e.g.,][]{Lilly2013}.
Thus, studies of MZR evolution can be used to constrain how the above processes evolve with cosmic time. 
Along these lines, \citet{Onodera2016} explain the evolution of the MZR normalization by suggesting that star-formation efficiency evolves weakly with redshift. 
However, \citet{Sanders2021} suggest an increasing star-formation efficiency with redshift to match their mild MZR evolution ($\mathrm{\Delta log(O/H)}/\mathrm{\Delta}z \simeq -0.1$).
An increase in star-formation efficiency with redshift is consistent with recent studies on the evolution of star-formation efficiency \citep[e.g.,][]{Liu2019}.
In this work, we also find a similar mild MZR evolution ($\mathrm{\Delta log(O/H)}/\mathrm{\Delta}z = -0.09\pm0.06$) at the low-mass end, which suggests that star-formation efficiency increases with redshift across a wide stellar mass range. 

The gas fraction of galaxies is another crucial factor for regulating the metallicity of galaxies.
Observations have found that the gas fraction increases significantly with redshift at fixed $M_\star$ and has been quantified by $\mu_\mathrm{gas} \equiv M_\mathrm{gas}/M_\star \propto (1+z)^{2.5}$ \citep[e.g.,][]{Tacconi2018}.
These large gas fractions can dilute the ISM of galaxies making lower gas-phase metallicity for higher redshift galaxies.
Theoretically, the evolution of the gas fraction is considered the driving force of MZR evolution in some cosmological simulations \citep[e.g.,][]{Ma2016, Torrey2019}.
However, the gas fraction is also predicted to increase with decreasing stellar mass at fixed redshift, which is quantified as $\mu_\mathrm{gas} \propto M_\star^{-0.4}$ for $z=2$ \citep[e.g.,][]{Torrey2019}.
In contrast to previous studies, our results indicate that the evolution rate of MZR appears to be the same in the low-mass and high-mass end, implying that the high gas fraction of dwarf galaxies at $z=2-3$ may have a weak effect on the evolution of MZR. 
In the future, direct measurements of the gas fraction in dwarf galaxies (e.g., using the Square Kilometre Array, SKA) are crucial to understanding the effect and contribution of gas reservoirs on galaxy metal enrichment.

\subsection{Existence of Fundamental Metallicity Relation in Dwarf Galaxies at $z=2-3$}

Previous works \citep[e.g.,][]{Ellison2008, Mannucci2010, Andrews2013} have suggested that a fundamental relation exists between metallicity and other 
galaxy properties.
This fundamental metallicity relation (FMR), correlates the metallicity with both stellar mass and SFR.
For a given stellar mass, the metallicity is enhanced with the decreasing SFR.
Several works probing large samples of galaxies from $z = 3$ to $z = 0$ have found little redshift evolution in the FMR in galaxies above 10$^{8}$\, M$_\odot$ \citep[e.g.,][]{Henry2013, Sanders2015, Sanders2018, Curti2020a, Henry2021}.
These results emphasize the dominance of stellar mass and SFR on the gas-phase metallicity of galaxies, regardless of the cosmic redshift.

The FMR relation can be explicitly parameterized using a two-dimensional projection with the relation between O/H and is written as
\begin{equation}
\mu_\alpha \equiv \operatorname{log} \left(\frac{M_\star}{\mathrm{M}_\odot}\right)-\alpha \times \operatorname{log} \left(\frac{\mathrm{SFR}}{\mathrm{M}_{\odot}~ \mathrm{yr}^{-1}}\right).
\end{equation}

In the above equation, $\alpha$ is introduced as the parameter to minimize the scatter in O/H.
For the dwarf galaxies at $z=2-3$ in our sample, the scatter of O/H is minimized at $\alpha_\mathrm{best} = 0.60\pm0.13$, suggesting the existence of FMR for $z=2-3$ dwarf galaxies. 

Based on our fitting, we construct an FMR with the linear function form of
\begin{equation}
    \mathrm{12+log(O/H)} = (0.20\pm 0.02) \mu_{\alpha_\mathrm{best}} + (6.59\pm 0.15),
\end{equation}
where $\mu_{\alpha_\mathrm{best}} = \operatorname{log} \left(M_\star/\mathrm{M}_\odot\right)-0.60\operatorname{log} \left(\mathrm{SFR}/{\mathrm{M}_{\odot}~ \mathrm{yr}^{-1}}\right)$.  We show the projection of the FMR, the O/H vs. $\mu_{\alpha_\mathrm{best}}$ {in Figure~\ref{fig:FMR}}.
As seen in Figure~\ref{fig:FMR}, we find that the FMR exists within a large stellar mass range ($M_\star = 10^{6}-10^{10} \mathrm{M}_\odot$) at $z=2-3$. 

The existence of the FMR in the low stellar mass end demonstrates that the SFR and stellar mass act as the main physical properties for determining the gas-phase metallicity in dwarf galaxies at $z=2-3$. 
However, it is important to note that the exact measurement of parameter $\alpha$ and interpretation of the FMR can vary depending on the method used to measure it.
For example, some works that have measured alpha for the FMR have found quite a range, with a systematic difference depending on whether the FMR was measured using a strong line calibration or direct metallicities in individual galaxies or stacks \citep[e.g.,][]{Andrews2013, Henry2021}.
In the context of these studies, the results presented in this paper suggest that the FMR for dwarf galaxies at $z=2-3$ follows a consistent $\alpha$ value with previous measurements \citep[$\alpha = 0.55-0.7$;][]{Andrews2013, Sanders2017, Sanders2021, Curti2020a}.
These measurements are over cosmic time ($z=0-3$), suggesting a slight evolution of FMR.
However, we note that the slope of FMR in the low-mass end is much shallower than previous measurements, and further studies will be needed to fully understand the evolution of the FMR over cosmic time.

In the future, deeper imaging and spectroscopic observations over larger survey area will allow us to probe the MZR and FMR relations out to even higher redshifts and down to even lower metallicity.
This will allow us to better understand the physical processes driving chemical enrichment at different epochs in the history of the Universe.

\section*{Acknowledgments}

{Our team thanks Feige Wang and Zuyi Chen for very helpful discussions.}
Z.C., M.L, X.L., Z.L., Y.W., \& S.Z. are supported by the National Key R\&D Program of China (grant no.\ 2018YFA0404503), the National Science Foundation of China (grant no.\ 12073014), the Tsinghua University Initiative Scientific Research Program (No.20223080023), and the science research grants from the China Manned Space Project with No. CMS-CSST2021-A05.
F.S.\ acknowledges support from the NRAO Student Observing Support (SOS) award SOSPA7-022. 
F.S.\ and E.E.\ acknowledge funding from JWST/NIRCam contract to the University of Arizona, NAS5-02105. 

This work is based on observations made with the NASA/ESA Hubble Space Telescope and  NASA/ESA/CSA James Webb Space Telescope. HST and JWST data were obtained from the Mikulski Archive for Space Telescopes at the Space Telescope Science Institute, which is operated by the Association of Universities for Research in Astronomy, Inc., under NASA contract NAS 5-03127 for JWST and NAS 5–26555 for HST. The JWST observations are associated with programs ERS-1324 and ERO-2736. The HST observations are associated with the Hubble Space Telescope Frontier Fields program and the Reionization Lensing Cluster Survey. The authors acknowledge the GLASS team for developing their observing program with a zero-exclusive-access period.
This work is based on data and catalog products from HFF-DeepSpace, funded by the National Science Foundation and Space Telescope Science Institute (operated by the Association of Universities for Research in Astronomy, Inc., under NASA contract NAS5-26555).
This work is based on observations taken by the RELICS Treasury Program (GO 14096) with the NASA/ESA HST, which is operated by the Association of Universities for Research in Astronomy, Inc., under NASA contract NAS5-26555.

{Some of the data presented in this paper were obtained from the Mikulski Archive for Space Telescopes (MAST) at the Space Telescope Science Institute. The JWST observations analyzed can be accessed via \dataset[doi:10.17909/91zv-yg35]{https://doi.org/10.17909/91zv-yg35} and
 \dataset[doi:10.17909/12rr-2a67]{https://doi.org/10.17909/12rr-2a67}.  The HFF data can be accessed via  \dataset[doi:10.17909/T9KK5N]{https://doi.org/10.17909/T9KK5N} and the RELICS data via \dataset[doi:10.17909/T9SP45]{https://doi.org/10.17909/T9SP45}.}

%

\vspace{5mm}
\facilities{JWST(NIRISS and NIRCam), HST(ACS and WFC3)}

\software{Astropy (\citealt{AstropyCollaboration2022}), Source Extractor (\citealt{Bertin1996}), Grizli (\citealt{Brammer2022}), BEAGLE (\citealt{Chevallard2016})}

\appendix
\section{Physical parameters for our sample}\label{sec:property}

In Table~\ref{tab:indi}, we show the measured and derived physical parameters of all {of} the galaxies in our sample, including galaxy ID defined by the \texttt{Grizli} spectrum extraction procedure, galaxy coordinates (R.A. and Decl.), galaxy redshift ($z_\mathrm{grism}$), stellar mass ($\mathrm{log}(M_\star/\mathrm{M}_\odot)$), dust attenuation ($A_V$), star formation rate (SFR) derived from Balmer emission lines, line ratios ($\mathrm{O_{32}}=\operatorname{log}$([\ion{O}{3}]/[\ion{O}{2}])) and ($\mathrm{O_{3}}=\operatorname{log}$([\ion{O}{3}]/H$\beta$)), and gas-phase ``metallicity (12+log(O/H)).
In Table~\ref{tab:line}, we present the observed line fluxes of [\ion{O}{2}], H$\beta$, [\ion{O}{3}], and H$\alpha$.
Note that these fluxes have not been corrected by dust attenuation and gravitational lensing.
The sample is divided into two redshift bins and presented for two fields A2744 and SMACS~0723.

\begin{deluxetable*}{ccccccccccc}
\tablenum{A1}
\tablecaption{Measured and {Derived} Properties of Individual Galaxies\label{tab:indi}}
\tablewidth{0pt}
\tablehead{
\colhead{ID} & \colhead{R.A.} & \colhead{Decl.} & \colhead{$z_{\rm grism}$} & \colhead{log$(M_\star/\mathrm{M}_\odot)$} & \colhead{$A_V$} & \colhead{SFR} & \colhead{$\mathrm{O}_{32}$} & \colhead{$\mathrm{O}_{3}$} &\colhead{$\rm 12+ log(O/H)$} & \colhead{$\mu$} \\
\colhead{} & \colhead{deg.} & \colhead{deg.} & \colhead{} & \colhead{} &
\colhead{} & \colhead{$\rm M_\odot~yr^{-1}$} & \colhead{} & \colhead{} & \colhead{} & \colhead{}
}
\decimalcolnumbers
\startdata
\multicolumn{11}{c}{\large A2744 Field}\\
\multicolumn{11}{c}{$1.8<z_\mathrm{grism}<2.3$}\\
   34 & 3.591921 & -30.415740 &      2.07 & $7.72^{+0.19}_{-0.14}$ & $0.33^{+0.05}_{-0.04}$ & $2.82\pm{0.15}$ & $0.87\pm{0.03}$ & $0.85\pm{0.02}$ & $8.03\pm{0.02}$ & 3.04 \\
  245 & 3.599965 & -30.409570 &      1.81 & $7.78^{+0.19}_{-0.17}$ & $0.39^{+0.06}_{-0.08}$ & $0.47\pm{0.05}$ & $1.60\pm{0.89}$ & $1.02\pm{0.11}$ & $7.60\pm{0.52}$ & 6.29 \\
  282 & 3.577672 & -30.408856 &      2.26 & $7.05^{+0.09}_{-0.09}$ & $0.01^{+0.01}_{-0.01}$ & $0.45\pm{0.06}$ & $0.69\pm{0.05}$ & $0.48\pm{0.03}$ & $8.13\pm{0.03}$ & 2.31 \\
  614 & 3.618857 & -30.403802 &      2.20 & $7.11^{+0.08}_{-0.08}$ & $0.19^{+0.04}_{-0.04}$ & $1.24\pm{0.25}$ & $0.86\pm{0.13}$ & $1.15\pm{0.28}$ & $8.03\pm{0.08}$ & 1.00 \\
  882 & 3.602338 & -30.400736 &      1.81 & $8.58^{+0.16}_{-0.15}$ & $0.20^{+0.08}_{-0.04}$ & $0.64\pm{0.08}$ & $0.04\pm{0.08}$ & $0.63\pm{0.08}$ & $8.52\pm{0.05}$ & 2.23 \\
 1018 & 3.606238 & -30.397984 &      2.28 & $6.89^{+0.24}_{-0.17}$ & $0.12^{+0.05}_{-0.03}$ & $1.05\pm{0.10}$ & $1.68\pm{0.61}$ & $0.80\pm{0.07}$ & $7.55\pm{0.36}$ & 1.78 \\
 1085 & 3.604176 & -30.397170 &      2.07 & $9.37^{+0.15}_{-0.20}$ & $0.38^{+0.05}_{-0.04}$ & $3.93\pm{0.26}$ & $0.16\pm{0.03}$ & $0.75\pm{0.09}$ & $8.44\pm{0.02}$ & 1.85 \\
 1333 & 3.606056 & -30.393533 &      2.18 & $8.42^{+0.18}_{-0.25}$ & $0.30^{+0.09}_{-0.07}$ & $1.48\pm{0.16}$ & $0.80\pm{0.06}$ & $1.13\pm{0.11}$ & $8.07\pm{0.03}$ & 1.68 \\
 1407 & 3.611638 & -30.392479 &      2.28 & $7.45^{+0.30}_{-0.30}$ & $0.13^{+0.04}_{-0.03}$ & $0.86\pm{0.08}$ & $1.03\pm{0.10}$ & $1.05\pm{0.09}$ & $7.93\pm{0.06}$ & 1.53 \\
 1409 & 3.599706 & -30.392378 &      1.87 & $7.85^{+0.15}_{-0.14}$ & $0.15^{+0.04}_{-0.04}$ & $0.71\pm{0.06}$ & $0.58\pm{0.24}$ & $0.63\pm{0.11}$ & $8.20\pm{0.14}$ & 1.95 \\
 1445 & 3.607556 & -30.391864 &      2.06 & $6.34^{+0.06}_{-0.04}$ & $0.16^{+0.06}_{-0.05}$ & $0.44\pm{0.13}$ & $1.85\pm{1.04}$ & $0.87\pm{0.08}$ & $7.45\pm{0.61}$ & 1.62 \\
 1461 & 3.604243 & -30.391652 &      1.88 & $8.06^{+0.14}_{-0.14}$ & $0.29^{+0.06}_{-0.07}$ & $0.85\pm{0.09}$ & $0.50\pm{0.08}$ & $0.92\pm{0.12}$ & $8.25\pm{0.05}$ & 1.71 \\
 1500 & 3.607616 & -30.391073 &      2.06 & $7.26^{+0.12}_{-0.12}$ & $0.47^{+0.06}_{-0.07}$ & $1.00\pm{0.13}$ & $0.71\pm{0.15}$ & $0.95\pm{0.14}$ & $8.12\pm{0.09}$ & 1.62 \\
 1503 & 3.603143 & -30.391055 &      2.18 & $7.57^{+0.15}_{-0.13}$ & $0.13^{+0.03}_{-0.03}$ & $1.72\pm{0.09}$ & $1.01\pm{0.05}$ & $0.87\pm{0.03}$ & $7.94\pm{0.03}$ & 1.80 \\
 1504 & 3.613264 & -30.391095 &      1.88 & $8.66^{+0.19}_{-0.23}$ & $0.42^{+0.06}_{-0.06}$ & $4.50\pm{0.32}$ & $0.48\pm{0.04}$ & $0.87\pm{0.04}$ & $8.26\pm{0.02}$ & 1.48 \\
 1573 & 3.609447 & -30.389859 &      2.21 & $8.96^{+0.12}_{-0.15}$ & $0.25^{+0.05}_{-0.05}$ & $2.09\pm{0.15}$ & $0.52\pm{0.05}$ & $0.66\pm{0.06}$ & $8.23\pm{0.03}$ & 1.57 \\
 1596 & 3.610005 & -30.389482 &      2.18 & $8.63^{+0.24}_{-0.31}$ & $0.32^{+0.07}_{-0.07}$ & $2.92\pm{0.24}$ & $0.52\pm{0.03}$ & $0.90\pm{0.05}$ & $8.23\pm{0.02}$ & 1.55 \\
 1685 & 3.603072 & -30.387831 &      1.88 & $6.23^{+0.10}_{-0.07}$ & $0.28^{+0.10}_{-0.08}$ & $0.29\pm{0.05}$ & $1.29\pm{0.72}$ & $1.07\pm{0.26}$ & $7.78\pm{0.43}$ & 1.73 \\
 1823 & 3.579059 & -30.385947 &      2.20 & $7.53^{+0.19}_{-0.16}$ & $0.14^{+0.03}_{-0.04}$ & $1.06\pm{0.06}$ & $0.81\pm{0.05}$ & $0.82\pm{0.04}$ & $8.06\pm{0.03}$ & 3.42 \\
 2032 & 3.592031 & -30.382505 &      1.87 & $6.87^{+0.11}_{-0.12}$ & $0.16^{+0.04}_{-0.03}$ & $0.56\pm{0.05}$ & $0.56\pm{0.06}$ & $0.77\pm{0.06}$ & $8.21\pm{0.04}$ & 2.43 \\
 2084 & 3.591101 & -30.381687 &      1.89 & $8.38^{+0.22}_{-0.13}$ & $0.27^{+0.02}_{-0.03}$ & $7.38\pm{0.22}$ & $0.70\pm{0.03}$ & $1.00\pm{0.02}$ & $8.13\pm{0.02}$ & 2.48 \\
 2152 & 3.585912 & -30.380660 &      1.87 & $7.32^{+0.32}_{-0.15}$ & $0.29^{+0.03}_{-0.03}$ & $2.07\pm{0.08}$ & $0.77\pm{0.04}$ & $0.86\pm{0.02}$ & $8.08\pm{0.02}$ & 2.82 \\
 2186 & 3.610360 & -30.380189 &      1.88 & $8.92^{+0.15}_{-0.15}$ & $0.48^{+0.05}_{-0.05}$ & $5.72\pm{0.33}$ & $0.05\pm{0.04}$ & $0.63\pm{0.08}$ & $8.51\pm{0.03}$ & 1.63 \\
 2303 & 3.607664 & -30.378280 &      1.85 & $6.38^{+0.21}_{-0.15}$ & $0.07^{+0.04}_{-0.03}$ & $0.19\pm{0.03}$ & $0.77\pm{0.30}$ & $0.43\pm{0.09}$ & $8.09\pm{0.18}$ & 1.76 \\
 2313 & 3.598602 & -30.378502 &      1.91 & $8.88^{+0.20}_{-0.25}$ & $0.27^{+0.07}_{-0.05}$ & $2.05\pm{0.17}$ & $0.45\pm{0.09}$ & $0.88\pm{0.11}$ & $8.27\pm{0.05}$ & 2.04 \\
 2330 & 3.599784 & -30.377873 &      1.89 & $7.96^{+0.22}_{-0.29}$ & $0.10^{+0.07}_{-0.05}$ & $0.51\pm{0.05}$ & $0.99\pm{0.19}$ & $0.74\pm{0.05}$ & $7.96\pm{0.11}$ & 2.03 \\
 2421 & 3.599383 & -30.375941 &      1.91 & $7.30^{+0.13}_{-0.12}$ & $0.06^{+0.02}_{-0.02}$ & $0.36\pm{0.03}$ & $1.05\pm{0.50}$ & $0.63\pm{0.10}$ & $7.92\pm{0.30}$ & 2.11 \\
\hline
\multicolumn{11}{c}{\large SMACS~0723 Field}\\
\multicolumn{11}{c}{$2.65<z_\mathrm{grism}<3.4$}\\
   50 & 110.821398 & -73.474602 &      2.98 & $8.78^{+0.07}_{-0.09}$ & $0.38^{+0.02}_{-0.04}$ &  $8.66\pm{1.42}$ &   - & $0.73\pm{0.07}$ & $8.35\pm{0.08}$ & 1.44 \\
 1347 & 110.823498 & -73.444440 &      3.37 & $8.22^{+0.06}_{-0.07}$ & $0.17^{+0.02}_{-0.01}$ & $14.00\pm{1.28}$ &   - & $0.89\pm{0.04}$ & $8.12\pm{0.09}$ & 1.73 \\
 1361 & 110.859711 & -73.444213 &      2.73 & $9.01^{+0.08}_{-0.07}$ & $0.12^{+0.03}_{-0.03}$ &  $6.27\pm{0.99}$ &   - & $0.97\pm{0.07}$ & $7.87\pm{0.15}$ & 1.79 \\
 1424 & 110.828726 & -73.442593 &      3.10 & $9.59^{+0.04}_{-0.04}$ & $0.31^{+0.01}_{-0.01}$ & $47.29\pm{1.63}$ &   - & $0.77\pm{0.01}$ & $8.30\pm{0.02}$ & 1.63 \\
 1475 & 110.863244 & -73.441489 &      2.73 & $8.33^{+0.09}_{-0.12}$ & $0.24^{+0.04}_{-0.07}$ &  $2.90\pm{0.51}$ &   - & $0.90\pm{0.07}$ & $8.10\pm{0.16}$ & 1.75 \\
 1491 & 110.859274 & -73.441230 &      2.73 & $9.83^{+0.02}_{-0.05}$ & $0.30^{+0.04}_{-0.02}$ &  $9.45\pm{1.18}$ &   - & $0.65\pm{0.05}$ & $8.43\pm{0.05}$ & 1.94 \\
 1520 & 110.850517 & -73.440461 &      3.04 & $7.71^{+0.10}_{-0.12}$ & $0.18^{+0.07}_{-0.05}$ &  $6.77\pm{0.90}$ &   - & $0.63\pm{0.05}$ & $8.44\pm{0.04}$ & 1.69 \\
\enddata
\tablecomments{To be continued.}
\end{deluxetable*}

\begin{deluxetable*}{ccccccccccc}
\tablenum{A1}
\tablecaption{(Continued)}
\tablewidth{0pt}
\tablehead{
\colhead{ID} & \colhead{R.A.} & \colhead{Decl.} & \colhead{$z_{\rm grism}$} & \colhead{log$(M_\star/\mathrm{M}_\odot)$} & \colhead{$A_V$} & \colhead{SFR} & \colhead{$\mathrm{O}_{32}$} & \colhead{$\mathrm{O}_{3}$} &\colhead{$\rm 12+ log(O/H)$} & \colhead{$\mu$} \\
\colhead{} & \colhead{deg.} & \colhead{deg.} & \colhead{} & \colhead{} &
\colhead{} & \colhead{$\rm \mathrm{M}_\odot~yr^{-1}$} & \colhead{} & \colhead{} & \colhead{} & \colhead{}
}
\decimalcolnumbers
\startdata
\multicolumn{11}{c}{\large A2744 Field}\\
\multicolumn{11}{c}{$2.65<z_\mathrm{grism}<3.4$}\\
  404 & 3.613460 & -30.406860 &      2.85 & $6.84^{+0.09}_{-0.05}$ & $0.06^{+0.03}_{-0.04}$ &  $2.56\pm{0.35}$ &  $0.90\pm{0.08}$ & $0.69\pm{0.06}$ & $8.01\pm{0.05}$ & 1.69 \\
  434 & 3.607437 & -30.406471 &      3.20 & $9.25^{+0.24}_{-0.42}$ & $0.35^{+0.05}_{-0.09}$ &  $6.79\pm{1.03}$ &  $0.52\pm{0.04}$ & $0.80\pm{0.05}$ & $8.23\pm{0.02}$ & 2.18 \\
  538 & 3.612998 & -30.405067 &      3.04 & $9.35^{+0.24}_{-0.43}$ & $0.43^{+0.04}_{-0.09}$ & $36.97\pm{3.36}$ &  $0.53\pm{0.03}$ & $0.46\pm{0.02}$ & $8.23\pm{0.01}$ & 1.69 \\
  903 & 3.609749 & -30.400342 &      2.97 & $9.68^{+0.13}_{-0.19}$ & $0.41^{+0.06}_{-0.04}$ &  $5.58\pm{1.85}$ & $-0.06\pm{0.04}$ & $0.77\pm{0.14}$ & $8.57\pm{0.02}$ & 1.73 \\
  988 & 3.613656 & -30.398646 &      2.85 & $9.29^{+0.28}_{-0.27}$ & $0.21^{+0.09}_{-0.04}$ &  $1.31\pm{1.02}$ &  $0.02\pm{0.07}$ & $0.87\pm{0.34}$ & $8.53\pm{0.04}$ & 1.56 \\
 1016 & 3.607800 & -30.398054 &      3.20 & $7.28^{+0.07}_{-0.06}$ & $0.06^{+0.02}_{-0.02}$ &  $3.10\pm{0.50}$ &  $0.83\pm{0.05}$ & $0.97\pm{0.07}$ & $8.05\pm{0.03}$ & 1.77 \\
 1042 & 3.610340 & -30.397893 &      3.36 & $7.79^{+0.03}_{-0.03}$ & $0.02^{+0.01}_{-0.01}$ & $14.97\pm{0.55}$ &  $0.96\pm{0.02}$ & $0.82\pm{0.02}$ & $7.98\pm{0.01}$ & 1.67 \\
 1206 & 3.607114 & -30.395622 &      2.98 & $8.27^{+0.31}_{-0.29}$ & $0.15^{+0.03}_{-0.03}$ &  $7.25\pm{0.46}$ &  $1.17\pm{0.05}$ & $0.85\pm{0.02}$ & $7.85\pm{0.03}$ & 1.72 \\
 1275 & 3.572769 & -30.394578 &      2.94 & $8.16^{+0.33}_{-0.37}$ & $0.24^{+0.04}_{-0.05}$ &  $1.58\pm{0.35}$ &  $0.67\pm{0.06}$ & $0.84\pm{0.09}$ & $8.15\pm{0.03}$ & 3.25 \\
 1394 & 3.611255 & -30.392440 &      2.98 & $7.13^{+0.16}_{-0.07}$ & $0.11^{+0.03}_{-0.03}$ &  $3.69\pm{0.35}$ &  $1.71\pm{0.62}$ & $0.44\pm{0.04}$ & $7.53\pm{0.37}$ & 1.57 \\
 1535 & 3.604198 & -30.390338 &      2.71 & $8.40^{+0.10}_{-0.16}$ & $0.09^{+0.06}_{-0.04}$ &  $1.07\pm{0.30}$ &  $0.67\pm{0.16}$ & $0.49\pm{0.13}$ & $8.15\pm{0.10}$ & 1.79 \\
 1692 & 3.604118 & -30.387855 &      2.84 & $7.52^{+0.25}_{-0.07}$ & $0.11^{+0.02}_{-0.02}$ &  $5.42\pm{0.35}$ &  $1.09\pm{0.04}$ & $0.92\pm{0.03}$ & $7.90\pm{0.02}$ & 1.76 \\
 2029 & 3.596739 & -30.382570 &      3.21 & $7.43^{+0.08}_{-0.06}$ & $0.20^{+0.03}_{-0.03}$ &  $5.89\pm{0.66}$ &  $1.30\pm{0.09}$ & $0.85\pm{0.05}$ & $7.77\pm{0.06}$ & 2.16 \\
 2060 & 3.585941 & -30.382101 &      3.06 & $8.52^{+0.32}_{-0.49}$ & $0.17^{+0.04}_{-0.04}$ &  $4.72\pm{0.35}$ &  $0.83\pm{0.02}$ & $0.90\pm{0.03}$ & $8.05\pm{0.01}$ & 3.33 \\
 2225 & 3.591399 & -30.379771 &      2.73 & $8.55^{+0.19}_{-0.15}$ & $0.70^{+0.04}_{-0.05}$ & $21.43\pm{1.34}$ &  $0.75\pm{0.01}$ & $0.91\pm{0.01}$ & $8.10\pm{0.01}$ & 2.50 \\
 2237 & 3.604111 & -30.379262 &      2.83 & $7.67^{+0.38}_{-0.39}$ & $0.22^{+0.06}_{-0.06}$ &  $1.42\pm{0.36}$ &  $0.79\pm{0.11}$ & $0.78\pm{0.11}$ & $8.08\pm{0.07}$ & 1.95 \\
 2287 & 3.607475 & -30.378498 &      2.69 & $8.34^{+0.11}_{-0.20}$ & $0.06^{+0.04}_{-0.03}$ &  $1.99\pm{0.25}$ &  $0.63\pm{0.08}$ & $0.46\pm{0.06}$ & $8.17\pm{0.05}$ & 1.83 \\
\enddata
\tablecomments{Column 1 is the source ID defined by our source detection procedure; Columns 2 and 3 are the equatorial coordinates right ascension (R.A.) and declination (Decl.) in equinox with an epoch of J2000; Column 4 is the redshift determined by NIRISS grism spectra; Column 5 is the stellar mass; Column 6 is the dust attenuation ($A_V$); Column 7 is the star formation rate derived from Balmer emission lines; Columns 8 and 9 are the line ratios O$_{32}=$ log([\ion{O}{3}]$\lambda\lambda4959,5007$/[\ion{O}{2}]$\lambda\lambda3727,3729$) and $\mathrm{O}_3$=log([\ion{O}{3}]$\lambda\lambda4959,5007$/H$\beta$); Column 10 is gas phase metallicity represented by oxygen abundance; Column 11 is the magnification of gravitational lensing effect.}
\end{deluxetable*}

\begin{deluxetable*}{ccccc}
\tablenum{A2}
\tablecaption{Measured emission line fluxes of Individual Galaxies\label{tab:line}}
\tablewidth{0pt}
\tablehead{
\colhead{ID} & \colhead{$f_{\rm [OII]}$} &  \colhead{$f_{\rm H\beta}$} & \colhead{$f_{\rm [O III]}$} & \colhead{$f_{\rm H\alpha}$} \\
\colhead{} & \colhead{$10^{-17}~\rm erg~s^{-1}~cm^{-2}$} & \colhead{$10^{-17}~\rm erg~s^{-1}~cm^{-2}$} & \colhead{$10^{-17}~\rm erg~s^{-1}~cm^{-2}$} & \colhead{$10^{-17}~\rm erg~s^{-1}~cm^{-2}$}
}
\decimalcolnumbers
\startdata
\multicolumn{5}{c}{\large A2744 Field}\\
\multicolumn{5}{c}{$1.8<z_\mathrm{grism}<2.3$}\\
   34 & $1.47\pm{0.10}$ & $1.54\pm{0.07}$ & $10.82\pm{0.10}$ &  $4.12\pm{0.11}$ \\
  245 & $0.08\pm{0.17}$ & $0.32\pm{0.08}$ &  $3.32\pm{0.09}$ &  $1.84\pm{0.12}$ \\
  282 & $0.42\pm{0.05}$ & $0.68\pm{0.04}$ &  $2.06\pm{0.06}$ &  $0.58\pm{0.08}$ \\
  614 & $0.23\pm{0.07}$ & $0.12\pm{0.07}$ &  $1.64\pm{0.09}$ &  $0.60\pm{0.12}$ \\
  882 & $2.64\pm{0.45}$ & $0.68\pm{0.12}$ &  $2.88\pm{0.12}$ &  $1.10\pm{0.11}$ \\
 1018 & $0.04\pm{0.05}$ & $0.29\pm{0.05}$ &  $1.85\pm{0.06}$ &  $0.89\pm{0.07}$ \\
 1085 & $2.42\pm{0.15}$ & $0.63\pm{0.13}$ &  $3.53\pm{0.15}$ &  $3.29\pm{0.13}$ \\
 1333 & $0.49\pm{0.06}$ & $0.23\pm{0.06}$ &  $3.10\pm{0.07}$ &  $1.09\pm{0.06}$ \\
 1407 & $0.23\pm{0.05}$ & $0.22\pm{0.05}$ &  $2.51\pm{0.06}$ &  $0.62\pm{0.05}$ \\
 1409 & $0.36\pm{0.20}$ & $0.32\pm{0.08}$ &  $1.38\pm{0.08}$ &  $1.05\pm{0.08}$ \\
 1445 & $0.02\pm{0.05}$ & $0.19\pm{0.04}$ &  $1.43\pm{0.05}$ &  $0.41\pm{0.12}$ \\
 1461 & $0.53\pm{0.09}$ & $0.20\pm{0.05}$ &  $1.65\pm{0.07}$ &  $0.92\pm{0.08}$ \\
 1500 & $0.29\pm{0.10}$ & $0.17\pm{0.05}$ &  $1.47\pm{0.07}$ &  $0.67\pm{0.08}$ \\
 1503 & $0.49\pm{0.05}$ & $0.68\pm{0.05}$ &  $5.00\pm{0.07}$ &  $1.63\pm{0.06}$ \\
 1504 & $1.68\pm{0.14}$ & $0.68\pm{0.06}$ &  $5.04\pm{0.08}$ &  $3.69\pm{0.07}$ \\
 1573 & $0.68\pm{0.08}$ & $0.49\pm{0.07}$ &  $2.25\pm{0.08}$ &  $1.47\pm{0.07}$ \\
 1596 & $1.25\pm{0.07}$ & $0.53\pm{0.06}$ &  $4.13\pm{0.09}$ &  $1.95\pm{0.07}$ \\
 1685 & $0.04\pm{0.06}$ & $0.06\pm{0.04}$ &  $0.73\pm{0.05}$ &  $0.33\pm{0.04}$ \\
 1823 & $0.75\pm{0.08}$ & $0.74\pm{0.07}$ &  $4.82\pm{0.09}$ &  $1.86\pm{0.08}$ \\
 2032 & $0.69\pm{0.10}$ & $0.42\pm{0.05}$ &  $2.50\pm{0.07}$ &  $1.01\pm{0.09}$ \\
 2084 & $5.90\pm{0.43}$ & $2.93\pm{0.14}$ & $29.62\pm{0.17}$ & $11.71\pm{0.14}$ \\
 2152 & $1.85\pm{0.16}$ & $1.51\pm{0.06}$ & $11.00\pm{0.10}$ &  $3.77\pm{0.08}$ \\
 2186 & $3.21\pm{0.29}$ & $0.85\pm{0.15}$ &  $3.63\pm{0.17}$ &  $4.86\pm{0.14}$ \\
 2303 & $0.11\pm{0.08}$ & $0.24\pm{0.04}$ &  $0.65\pm{0.05}$ &  $0.27\pm{0.04}$ \\
 2313 & $1.51\pm{0.31}$ & $0.57\pm{0.14}$ &  $4.29\pm{0.16}$ &  $2.61\pm{0.14}$ \\
 2330 & $0.22\pm{0.10}$ & $0.39\pm{0.05}$ &  $2.13\pm{0.06}$ &  $0.80\pm{0.05}$ \\
 2421 & $0.10\pm{0.11}$ & $0.26\pm{0.06}$ &  $1.10\pm{0.07}$ &  $0.60\pm{0.05}$ \\
\hline
\multicolumn{5}{c}{\large SMACS~0723 Field}\\
\multicolumn{5}{c}{$2.65<z_\mathrm{grism}<3.4$}\\
   50 &   - & $0.77\pm{0.12}$ &  $4.12\pm{0.18}$ &  - \\
 1347 &   - & $1.43\pm{0.13}$ & $11.19\pm{0.20}$ &  - \\
 1361 &   - & $1.18\pm{0.18}$ & $11.05\pm{0.21}$ &  - \\
 1424 &   - & $4.72\pm{0.14}$ & $27.79\pm{0.21}$ &  - \\
 1475 &   - & $0.46\pm{0.08}$ &  $3.65\pm{0.11}$ &  - \\
 1491 &   - & $1.54\pm{0.18}$ &  $6.86\pm{0.22}$ &  - \\
 1520 &   - & $0.86\pm{0.09}$ &  $3.68\pm{0.13}$ &  - \\
\enddata
\tablecomments{To be continued.}
\end{deluxetable*}

\begin{deluxetable*}{ccccc}
\tablenum{A2}
\tablecaption{(Continued)}
\tablewidth{0pt}
\tablehead{
\colhead{ID} & \colhead{$f_{\rm [OII]}$} &  \colhead{$f_{\rm H\beta}$} & \colhead{$f_{\rm [O III]}$} & \colhead{$f_{\rm H\alpha}$} \\
\colhead{} & \colhead{$10^{-17}~\rm erg~s^{-1}~cm^{-2}$} & \colhead{$10^{-17}~\rm erg~s^{-1}~cm^{-2}$} & \colhead{$10^{-17}~\rm erg~s^{-1}~cm^{-2}$} & \colhead{$10^{-17}~\rm erg~s^{-1}~cm^{-2}$}
}
\decimalcolnumbers
\startdata
\multicolumn{5}{c}{\large A2744 Field}\\
\multicolumn{5}{c}{$2.65<z_\mathrm{grism}<3.4$}\\
  404 & $0.27\pm{0.05}$ & $0.44\pm{0.06}$ &  $2.18\pm{0.08}$ &  - \\
  434 & $1.51\pm{0.13}$ & $0.80\pm{0.10}$ &  $5.03\pm{0.11}$ &  - \\
  538 & $2.91\pm{0.16}$ & $3.44\pm{0.14}$ &  $9.83\pm{0.18}$ &  - \\
  903 & $3.89\pm{0.29}$ & $0.58\pm{0.19}$ &  $3.41\pm{0.20}$ &  - \\
  988 & $1.22\pm{0.14}$ & $0.17\pm{0.13}$ &  $1.28\pm{0.13}$ &  - \\
 1016 & $0.59\pm{0.06}$ & $0.43\pm{0.07}$ &  $3.97\pm{0.08}$ &  - \\
 1042 & $1.34\pm{0.06}$ & $1.82\pm{0.07}$ & $12.14\pm{0.11}$ &  - \\
 1206 & $0.49\pm{0.05}$ & $1.03\pm{0.05}$ &  $7.24\pm{0.08}$ &  - \\
 1275 & $0.58\pm{0.07}$ & $0.39\pm{0.08}$ &  $2.68\pm{0.11}$ &  - \\
 1394 & $0.03\pm{0.04}$ & $0.50\pm{0.04}$ &  $1.38\pm{0.06}$ &  - \\
 1535 & $0.14\pm{0.05}$ & $0.21\pm{0.06}$ &  $0.65\pm{0.07}$ &  - \\
 1692 & $0.62\pm{0.06}$ & $0.93\pm{0.05}$ &  $7.65\pm{0.08}$ &  - \\
 2029 & $0.29\pm{0.06}$ & $0.82\pm{0.09}$ &  $5.80\pm{0.10}$ &  - \\
 2060 & $1.37\pm{0.07}$ & $1.19\pm{0.07}$ &  $9.38\pm{0.10}$ &  - \\
 2225 & $3.97\pm{0.08}$ & $2.74\pm{0.08}$ & $22.15\pm{0.13}$ &  - \\
 2237 & $0.23\pm{0.06}$ & $0.24\pm{0.06}$ &  $1.42\pm{0.07}$ &  - \\
 2287 & $0.30\pm{0.05}$ & $0.43\pm{0.05}$ &  $1.25\pm{0.06}$ &  - \\
\enddata
\tablecomments{Column 1 is the source ID defined by our source detection procedure; Columns 2-5 present the observed emission line fluxes of [\ion{O}{2}], H$\beta$, [\ion{O}{3}], and H$\alpha$. Note that the ``-'' value means that the grism observations do not cover the corresponding wavelength.}
\end{deluxetable*}


\bibliography{sample631}{}
\bibliographystyle{aasjournal}



\end{document}